%% file: ms.tex
\documentclass[a4paper,11pt]{article}
\pdfoutput=1 

\usepackage[utf8]{inputenc}

\usepackage{jinstpub} 

\usepackage[greek,american]{babel}
\renewcommand{\u}{\greektext m\latintext}

\renewcommand{\emph}[1]{\textbf{#1}}

\title{Novel event classification based on spectral analysis of scintillation waveforms in Double Chooz}

\input{DCAuthors.tex}
\collaboration{Double Chooz collaboration}

\abstract{Liquid scintillators are a common choice for neutrino physics experiments, but their capabilities to perform background rejection by scintillation pulse shape discrimination is generally limited in large detectors. This paper describes a novel approach for a pulse shape based event classification developed in the context of the Double Chooz reactor antineutrino experiment. Unlike previous implementations, this method uses the Fourier power spectra of the scintillation pulse shapes to obtain event-wise information. A classification variable built from spectral information was able to achieve an unprecedented performance, despite the lack of optimization at the detector design level. Several examples of event classification are provided, ranging from differentiation between the detector volumes and an efficient rejection of instrumental light noise, to some sensitivity to the particle type, such as stopping muons, ortho-positronium formation, alpha particles as well as electrons and positrons. In combination with other techniques the method is expected to allow for a versatile and more efficient background rejection in the future, especially if detector optimization is taken into account at the design level.}

\arxivnumber{1710.04315}

\begin{document}
\maketitle

\flushbottom

\section{Introduction}
Liquid scintillators have a long tradition in neutrino physics, ever since the discovery of the neutrino itself. They are a popular choice for the detector material, as they can be produced at reasonable costs and in the large quantities that are necessary for neutrino detection. While water Cherenkov detectors are a viable cost-effective alternative at high neutrino energies, scintillators have the additional advantage of high light yield, which makes them also suitable for the detection of neutrinos in the MeV range, as is the case for antineutrinos from nuclear reactors. Furthermore, the source materials used in the production can be purified with chemical and physical methods, so liquid scintillators can be created with high optical and radiochemical purity, which is an important aspect for the control of radioactive backgrounds.

An important downside, however, is the limited particle identification capability. The intensity of the scintillation light is usually the only directly observable quantity, and only the energy deposited in the interaction can be deduced from this information. In principle, the shape of the scintillation pulse can provide information about the type of particle in the event, since different particles have different energy deposition mechanisms and thus excite the scintillator in different ways. This method is known as pulse shape discrimination (PSD) and is successfully used in compact or segmented detectors. But they usually do not work well in large-scale detectors, since the original scintillation pulse shapes are ``smeared out'' and distorted over the large distances.

In this paper we demonstrate the capabilities and prospects of a novel approach to particle identification (PID) in large liquid scintillation detectors. It was studied in the context of the Double Chooz experiment and yielded a performance superior to classical pulse shape discrimination~\cite{Wagner}. In this procedure the Fourier spectrum of a scintillation pulse is analyzed, rather than the shape of the pulse itself. We call this approach \textit{spectral shape discrimination} (SSD), to distinguish it from the more traditional \textit{pulse shape discrimination} (PSD) techniques.\\
\\
After a short introduction about the Double Chooz experiment in Section~\ref{sec:DCexp}, the spectral shape discrimination approach is described in Section~\ref{approach}, where a classifier variable is created from the Fourier spectra of the scintillation pulse shapes. Section~\ref{sec:performance} then investigates the performance of this classifier for three different tasks: the separation of different scintillators used in the detector, the rejection of so-called \textit{light noise} background, and particle identification, all by means of the recorded pulse shapes of the respective events. This is followed by a discussion of current limitations of the technique and possible solutions in Section~\ref{sec:limitations}. The paper concludes with a summary in Section~\ref{sec:conclusions}.

\section{The Double Chooz experiment}\label{sec:DCexp}
Double Chooz is a reactor antineutrino experiment located at the Chooz nuclear power plant in northern France. It determines the mixing angle $\theta_{13}$ of the leptonic mixing matrix via the observation of electron-antineutrino disappearance at short baseline. The experiment consists of two identically constructed liquid scintillator antineutrino detectors at mean distances of 400~m and 1050~m from the reactor cores.

The scintillators are based on organic hydrocarbons, which also provide the free protons as neutrino targets. The antineutrinos from the reactor are detected via the inverse beta decay (IBD) reaction
		$\overline{\nu}_e + p \rightarrow n + e^+$,
in which an electron antineutrino transforms a Hydrogen nucleus into a neutron and a positron. The positron quickly deposits its energy and subsequently annihilates, giving a \textit{prompt} signal, while the neutron first thermalizes and is then captured by a Gadolinium nucleus. Upon deexcitation the Gd nucleus releases several gammas with a total energy of $\approx8$~MeV, producing a \textit{delayed} signal~\cite{DC3}. Alternatively, neutron capture can occur on a Hydrogen nucleus, which upon deexcitation emits a single gamma of $\approx2.2$~MeV~\cite{H2}.

The Double Chooz detectors are described in detail in Ref.~\cite{DC2}. They are divided into four concentric volumes, each one containing different organic liquids:
		\begin{description}
		\item[Neutrino Target (NT)] The innermost volume is an acrylic vessel containing 10.3~m$^3$ of liquid scintillator mixture of
		80\% n-dodecane and 20\% PXE, with 7~g/l PPO and 20~mg/l bis-MSB as fluors~\cite{Production}. The scintillator is doped
		with	1~g/l of Gadolinium in form of a soluble organic metal complex Gd(thd)$_3$ to capture neutrons from an antineutrino
		interaction.
		\item[Gamma Catcher (GC)] The Target is surrounded by the GC vessel, which contains 22.3~m$^3$ of a scintillator mixture
		of 46\% n-dodecane, 50\% mineral oil and 4\% PXE, as well as the fluors PPO and bis-MSB, but without Gd
		doping~\cite{Production}. Its purpose is to absorb gammas escaping from the Target volume and convert them into visible
		light, so that they can be detected by the PMT system.
		\item[Buffer] The Buffer volume is a stainless steel vessel with 390 PMTs (Hamamatsu R7081, 10-inch) to detect scintillation
		light originating from the inner detector volumes. The steel tank encloses a non-scintillating mineral oil that shields the inner
		volumes from external radioactivity.
		\item[Inner Veto (IV)] The outermost scintillator volume is the Inner Veto, which is optically separated from the other
		volumes.	It is equipped with 78 PMTs for photon detection and serves as an active veto to reject cosmic muons and external
		radioactivity. 
		\end{description}
The PMT signals are recorded by 8-bit FADC electronics at 500~MHz~\cite{Electronics}, which allows for an accurate digitization of the pulses. For each event the digitized waveforms are recorded individually for each PMT. The availability of high-quality waveforms is essential for the analysis technique presented in this paper. Additionally, an Outer Veto system made of several plastic scintillator panels is installed above the detector and used for muon tracking.\\
\\
The chemical compositions of the Target and Gamma Catcher scintillators lead to characteristic shapes of the scintillation pulse. Especially the different fluor concentrations govern the effectiveness of the energy transfer mechanism and determine the pulse shapes. In the Target the presence of the Gd-complex further affects the energy transfer and leads to shorter pulses. The type of particle also influences the shape of the resulting pulse, but to a lesser extent than the chemical composition. The energy loss function $\left<\mathrm{d}E/\mathrm{d}x\right>$ is characteristic for each particle and influences the ratio of molecules that are excited into singlet and triplet states. These states deexcite with different time constants, leading to different shapes of the scintillation pulse from which the particle species can be determined~\cite{Aberle,Wagner}.

The novel spectral shape discrimination technique described in this document attempts to exploit these pulse shape differences to gain information about the type of scintillator in which an event took place, as well as to infer the type of particle in an interaction.

\section{The Spectral Shape Discrimination (SSD) approach}\label{approach}
Spectral shape discrimination aims at a discrimination between different event classes in Fourier space, making use of the possibility that the frequency domain representation may reveal characteristics about an event that were not or less clearly visible in the time domain.\\
\\
The FADC signals are sampled at $k$ equidistant points in the time domain, producing a function $p(t_n)$ at discrete values of $t_n$. The signal is only recorded over a finite time window of length $T$, so $t_n = n\,T/k$ with $n = 0, \dots, k$. This digital signal can be transformed into the frequency domain with help of the discrete Fourier transform. It creates $k$ complex Fourier coefficients $\mathcal{F}_j$ given by
		\begin{equation}
		\mathcal{F}_j \{ p \} = {\sum}_{n=0}^{k-1} \, p(t_n)\, e^{-2\pi i \; j\, n / k}.
		\end{equation}
In the case of purely real-valued input signals, the discrete Fourier transform produces $k/2+1$ independent Fourier coefficients. Each Fourier coefficient $\mathcal{F}_j$ can be separated into a modulus and a phase $\phi_j$ via $\mathcal{F}_j = s_j \, e^{i \phi_j}$, where the modulus
		\begin{equation}
		s_j \coloneqq \left| \mathcal{F}_j \{ p \} \right|
		\end{equation}
represents the power of the frequency $j$ in the original signal.\\
\\
The Fourier transform gives a complete description of the original pulse and has several very desirable properties compared to the original signal. For example, the Fourier spectrum is independent of the position of the pulse in the readout window. A shift of the whole pulse along the time axis only adds an imaginary phase factor to all the Fourier components. The spectrum, which is built from the magnitude of the complex Fourier coefficients, is left unchanged. This also means that parameters like pulse start time or peak time do not have to be calculated, removing a potential source of uncertainty. In addition, while the power spectrum of noise extends into the high-frequency region, the signals themselves are mainly composed of lower frequencies, so part of the noise contribution is separated from the signal.\\
\\
For the identification and separation of different event classes in Double Chooz we first created the Fourier power spectrum, then a discrimination parameter is designed from the spectral coefficients. This parameter is a measure of the shape of the pulses and can be used to distinguish different classes of events. The power spectrum is created offline from recorded FADC data in the following way:
		\begin{figure}[t]
		\centering
		\includegraphics[width=0.476\textwidth]{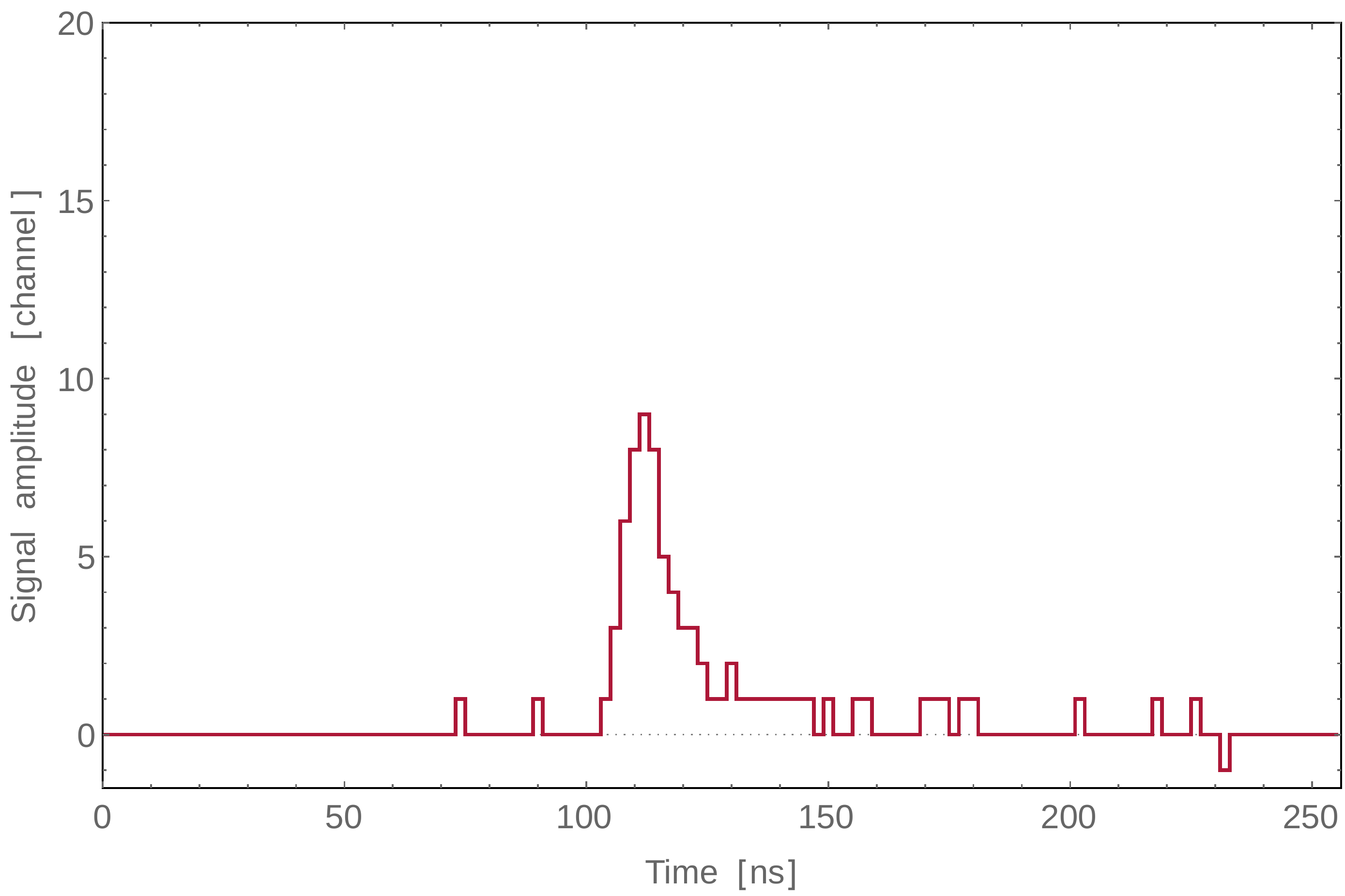}
		\hfill
		\includegraphics[width=0.49\textwidth]{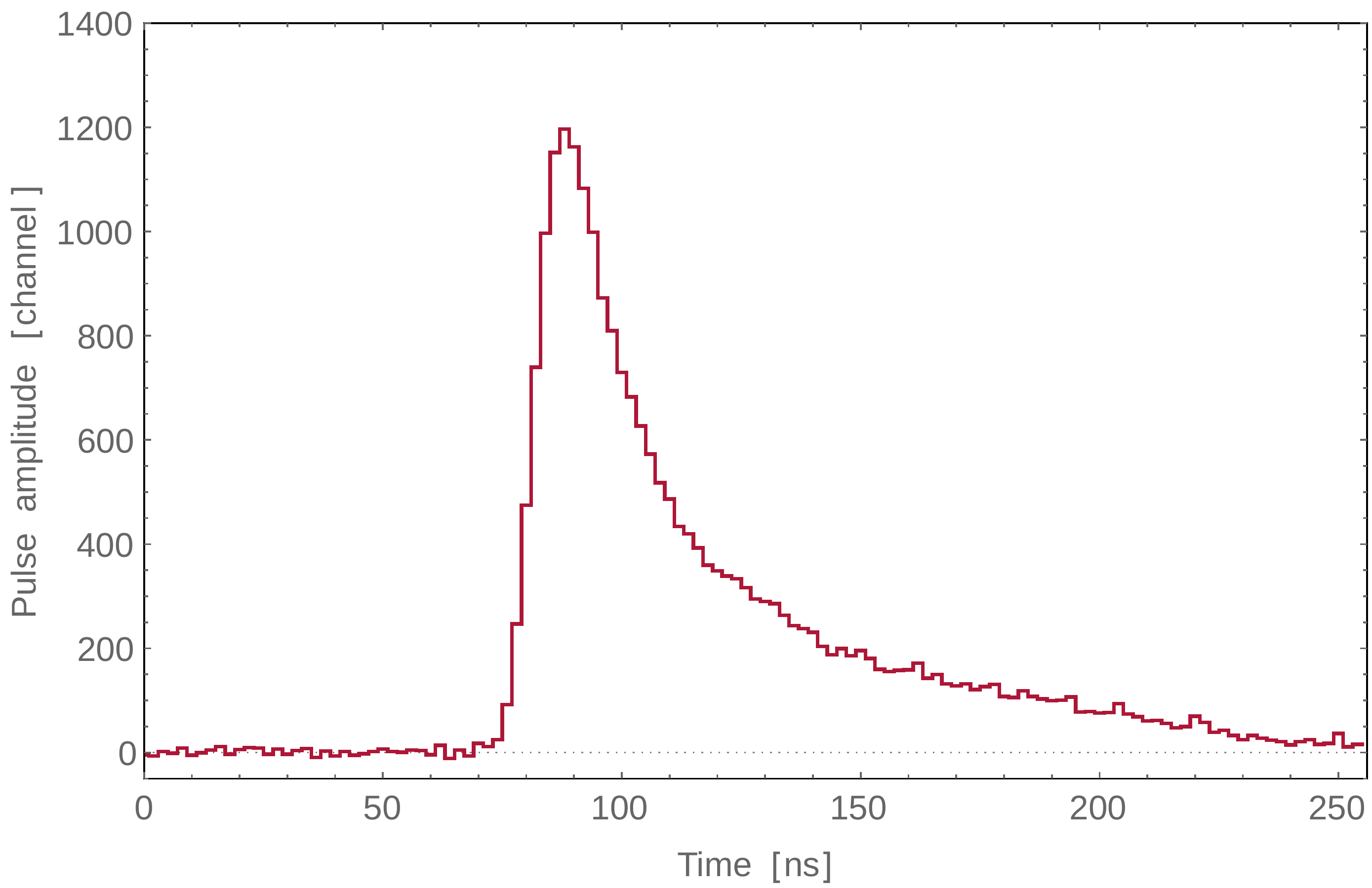}
		\caption{
		\emph{(Left)} Example of a single-photoelectron signal recorded by a PMT. The baseline was already subtracted and the pulse was converted into a positive signal.
\emph{(Right)} Example of a reconstructed scintillation pulse shape after summation of the 390 time-corrected individual PMT signals (before normalization). The displayed event deposited an energy of approximately 3~MeV.\label{img:pulses}}
		\end{figure}
		\begin{enumerate}
		\item First, the scintillation pulse shape of the event has to be reconstructed from the individual PMT signals. The left panel of Figure~\ref{img:pulses} shows a digitized single-photoelectron (SPE) signal recorded by a single PMT. The size and shape of SPE signals can vary, but they are large enough to be clearly separated from random fluctuations. The average noise level of the baseline is less than one channel in amplitude, while even the smallest SPE signals are at least three channels above the baseline. If a PMT was hit by a photon can thus be checked by a threshold algorithm. If yes, the recorded information is retrieved, the baseline is subtracted from the signal. For convenience, the signals are also converted to positive pulses. PMTs which did not detect a photon are disregarded in order to avoid picking up unnecessary noise from the baseline.
		\item Then the individual signals from each PMT must be time corrected. For this purpose the
		distance between the reconstructed event vertex and each PMT position is calculated. Then the photon time of flight is
		determined taking into account the refractive index in the liquids. A linear light path and a constant refractive index of the
		medium are assumed. In addition each recorded pulse has to be corrected for the time offset T$_0$ of the respective PMT,
		which is known from calibration runs with the Inner Detector Light Injection system~\cite{DC3}.
		Both corrections are applied to the pulses by shifting them by a corresponding amount of time in the readout window.
		\item All T$_0$- and time of flight-corrected waveforms are summed up. An example of a reconstructed pulse is shown in the right panel of Figure~\ref{img:pulses}. Since the analysis is to be independent of the event energy or pulse size, the sum pulse is then normalized to integral one. Otherwise the spectrum would also contain energy information. The integral over the corrected sum pulse is calculated and the pulse is divided by this value. To first order, the resulting normalized spectrum will be independent of the energy.
		\item The normalized sum pulse $p(t)$ is then transformed into the frequency domain with a Fast Discrete Fourier Transform
		(FDFT) algorithm. This produces the complex-valued Fourier coefficients
		\mbox{$\omega_j \coloneqq \mathcal{F}_j \{ p \}$} of the signal $p$. In Double Chooz there are 128 real-valued ADC samples,
		which produce 64 independent complex-valued Fourier coefficients $\omega_j$, with \mbox{$j=1,\dots,64$}, plus a normalization
		constant $\omega_0$. The power spectrum is then calculated as the absolute values of these coefficients:
		\mbox{$s_j = \sqrt{\omega_j\,\omega_j^\star}$}
		\end{enumerate}
		\begin{figure}[tbp]
		\centering
		\includegraphics[width=0.48\textwidth]{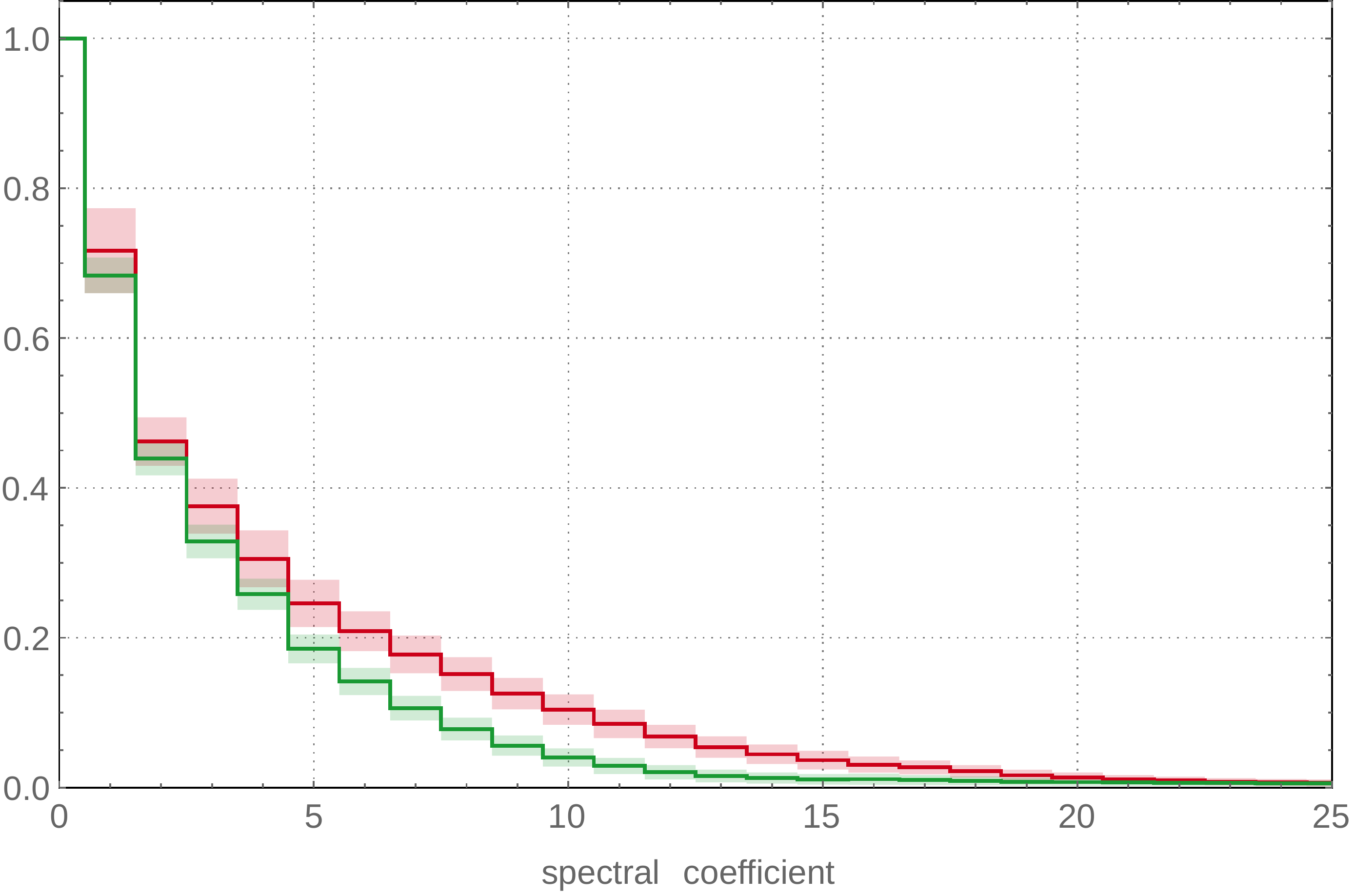}
		\hfill
		\includegraphics[width=0.48\textwidth]{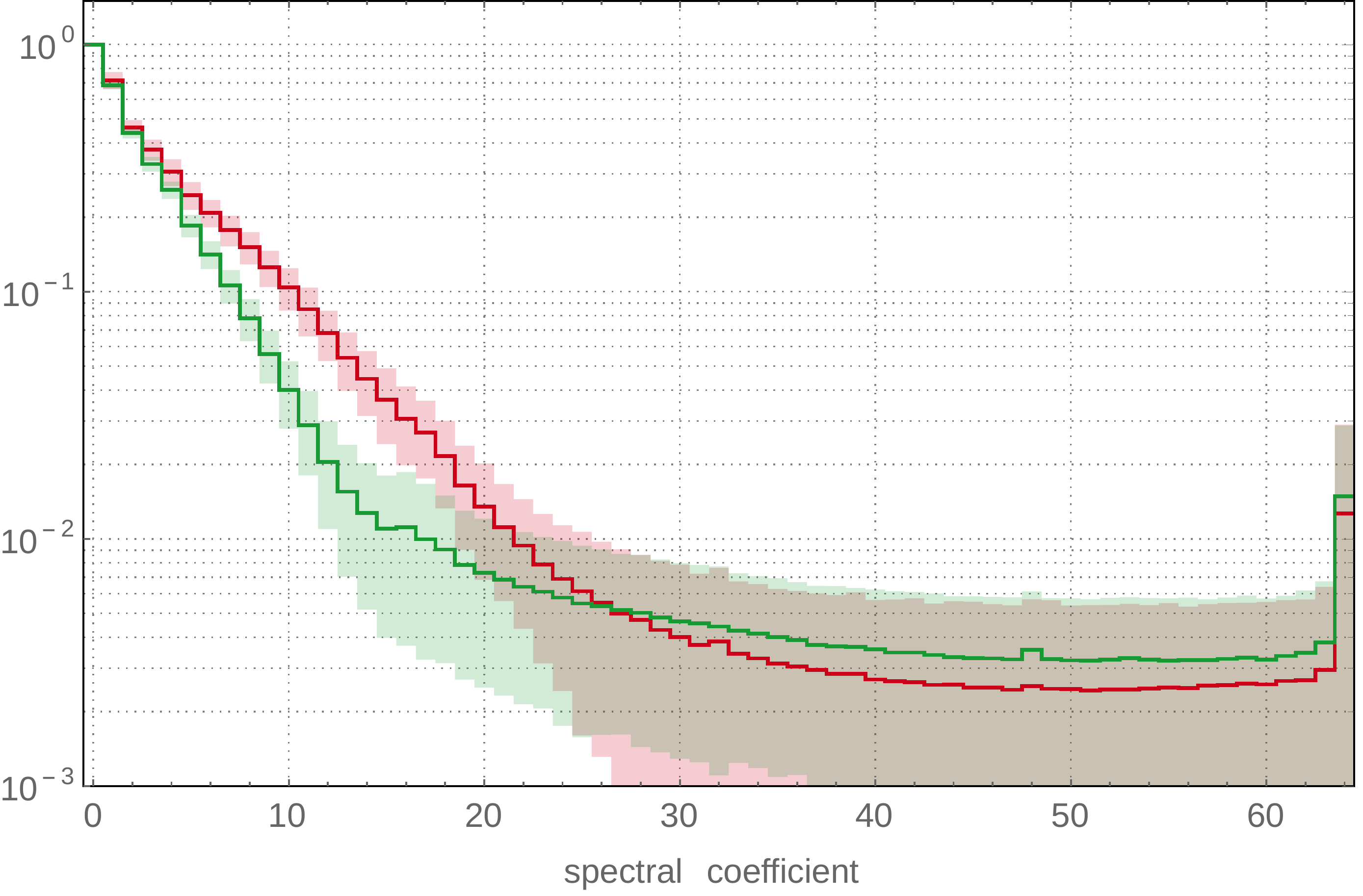}
		\caption{\emph{(Left)} Average power spectra for different classes of events. The plot shows the first 20 spectral
		coefficients of physics events in the Target (red) and Gamma Catcher (green) volumes obtained with a $^{252}$Cf calibration
		source in the respective volumes.\label{img:spectrum}
		\emph{(Right)} The same spectra in logarithmic scale over the whole range of coefficients. Above coefficient 20
		the two spectra overlap almost completely, so that there is no discrimination power in the spectral coefficients any more.
		The spike at coefficients 64 is due to an additive high-frequency electronic noise induced in the FADC by the
		clock~\cite{Electronics}, the smaller peaks at coefficients 48 and 32 are its respective subharmonics.}
		\end{figure}
In the following a classifier is created from the Fourier power spectra. Figure~\ref{img:spectrum} shows the average spectra for physics events in the Target and in the Gamma Catcher volumes in linear and logarithmic scale. The events were obtained from calibration runs with a $^{252}$Cf source in each of the two volumes. The Target has shorter scintillation pulses and exhibits a wider power spectrum. The Gamma Catcher, on the other hand, has longer scintillation pulses in the time domain, and consequently a slightly narrower Fourier spectrum.\\
\\
To characterize the spectra of the different event classes, we now construct a classification parameter from the spectra. As seen in Figure~\ref{img:spectrum} the curves can be easily distinguished by their integrals, so an intuitive approach is to sum over all the Fourier coefficients in the signal range. A first discriminator variable can be constructed as
		\begin{equation}
		\Omega_0 \coloneqq \sum_{j=1}^{n} \left| \mathcal{F}_j \{ p(t) \} \right|
		\label{eq:Omega}
		\end{equation}
where $n$ is the number of Fourier coefficients which are considered. The real detector signals contain a certain amount of noise, which is also transformed into the Fourier domain and dominate in the higher frequencies. The cutoff $n$ should be chosen such that the noise-dominated coefficients are not taken into account, as they would mainly add statistical fluctuations without contributing to the separation capability. In Figure~\ref{img:spectrum} the transition into the noise-dominated regime happens between the coefficients 20 and 30, where the exponential-like curves start to flatten out, therefore $n=20$ was chosen.\footnote{As long as $n$ does not cover noise-dominated coefficients, the exact choice of $n$ does not have a significant effect on the results, due to the weight that will be applied to the spectral coefficients later on.} The lower limit of the summation starts at one, since $\omega_0$ is always 1 (because of the normalization of the pulse) and does not contain any discrimination power.\\
\\
The parameter defined in equation~\eqref{eq:Omega} can already be used for classification of event categories. However, when pre-categorized data is available, e.g. from a calibration with sources, the parameter can be generalized and improved with weights $w_j$ applied to the spectral coefficients:
		\begin{equation}
		\Omega \coloneqq \sum_{j=1}^{n} w_j \left| \mathcal{F}_j \{ p(t) \} \right|
		\label{eq:OmegaOpt}
		\end{equation}
Certain coefficients may carry more information about the event category than others, and are more useful for the discrimination of different event categories. In Figure~\ref{img:spectrum} it is seen that both curves are well separated between coefficients 5 to 15, where their error bands do not overlap. It makes sense to give such coefficients more importance for the purpose of separation of Gamma Catcher and Target events. Coefficients with a larger overlap should in turn receive a lower weight, or be disregarded completely when the overlap becomes too large.

The specific form of these weights can be defined in various ways, depending on the choice of the optimization criterion. Here, the best separation is defined as the one that minimizes the overlap between the two distributions of $\Omega$ values. Under this premise we consider two classes of events $A$ and $B$ with different power spectra. Suppose that each spectral coefficient $s_j^A$ of class $A$ follows a normal distribution $\mathcal{N}(\mu_j^A,\sigma_j^A)$ with mean $\mu_j^A$ and standard deviation $\sigma_j^A$. Then the resulting parameter $\Omega_A$ has a mean value
		\begin{equation}
		\mu_A = \sum_j w_j \, \mu_j^A
		\end{equation}
and standard deviation
		\begin{equation}
		\sigma_A = \sqrt{ \sum_j \left( w_j \, \sigma_j^A \right)^2}
		\end{equation}
for events of class $A$, and likewise for events of class $B$.
Without loss of generality we can assume that $\mu_A < \mu_B$. Then the area of overlap $S$ of the two Gaussian probability distributions for $\Omega_A$ and $ \Omega_B$ is
		\begin{equation}
		S = 1-\mathrm{Erf}\left( \frac{x-\mu_A}{\sigma_A \sqrt{2}} \right)
		+\mathrm{Erf}\left( \frac{x-\mu_B}{\sigma_B \sqrt{2}} \right)
		\label{area}
		\end{equation}
as a function of the weights $w_j$. $\mathrm{Erf}$ is the error function and
		\begin{equation}
		x=\frac{\mu_B^{} \sigma_A^2 - \mu_A^{} \sigma_B^2 + \sigma_A \sigma_B \sqrt{\Delta\mu^2
		+ 2\, \Delta\sigma^2 \log(\sigma_A/\sigma_B)}}{\sigma_A^2 - \sigma_B^2}
		\end{equation}
is the intersection point between the two curves, where $\Delta \sigma^2 =\sigma^2_B-\sigma^2_A$ and $\Delta \mu=\mu_B-\mu_A$. Equation~\ref{area} can be minimized analytically over all weights $w_j$. The resulting set of weights $w_j$ then represent the best possible weights for the spectral coefficients under the criterion of minimal overlap between the distributions.\footnote{The weights obtained from the minimization procedure are only unique up to a common scale factor, which can be chosen freely without affecting the results. In this paper the scale factor was chosen such that all relevant values of $\Omega$ lie in a range from 0 to 10.}\\
\\
In this particular case the parameter was optimized for the separation of Gamma Catcher and Target events (according to the spectra in Figure~\ref{img:spectrum}), but the procedure can be performed for any two classes of events for which reference spectra exist. This way, dedicated classifier variables can be designed for the separation of two different classes of events, for instance physics events and light noise (Section~\ref{sec:ln}) or alpha particles and electrons (Section~\ref{sec:PID}). Each task produces another set of weights and the resulting optimized classifiers will assume different values.

The weighting procedure described above essentially produces the best linear combination of the spectral coefficients for the purpose of separating two classes of events. Nonlinear methods could achieve an even better separation than the linear combination and might be employed for a possibly increased performance (see Section~\ref{sec:limitations}).

\section{Performance of the discriminator}\label{sec:performance}

\subsection{GC/Target separation}\label{sec:GCTargetSeparation}
The performance of this method is first evaluated with physics events occurring in the Target or Gamma Catcher volume. Due to their different chemical compositions the two scintillators produce rather different pulse shapes in response to the same particle.

		\begin{figure}[t]
		\centering
		\includegraphics[width=0.48\textwidth]{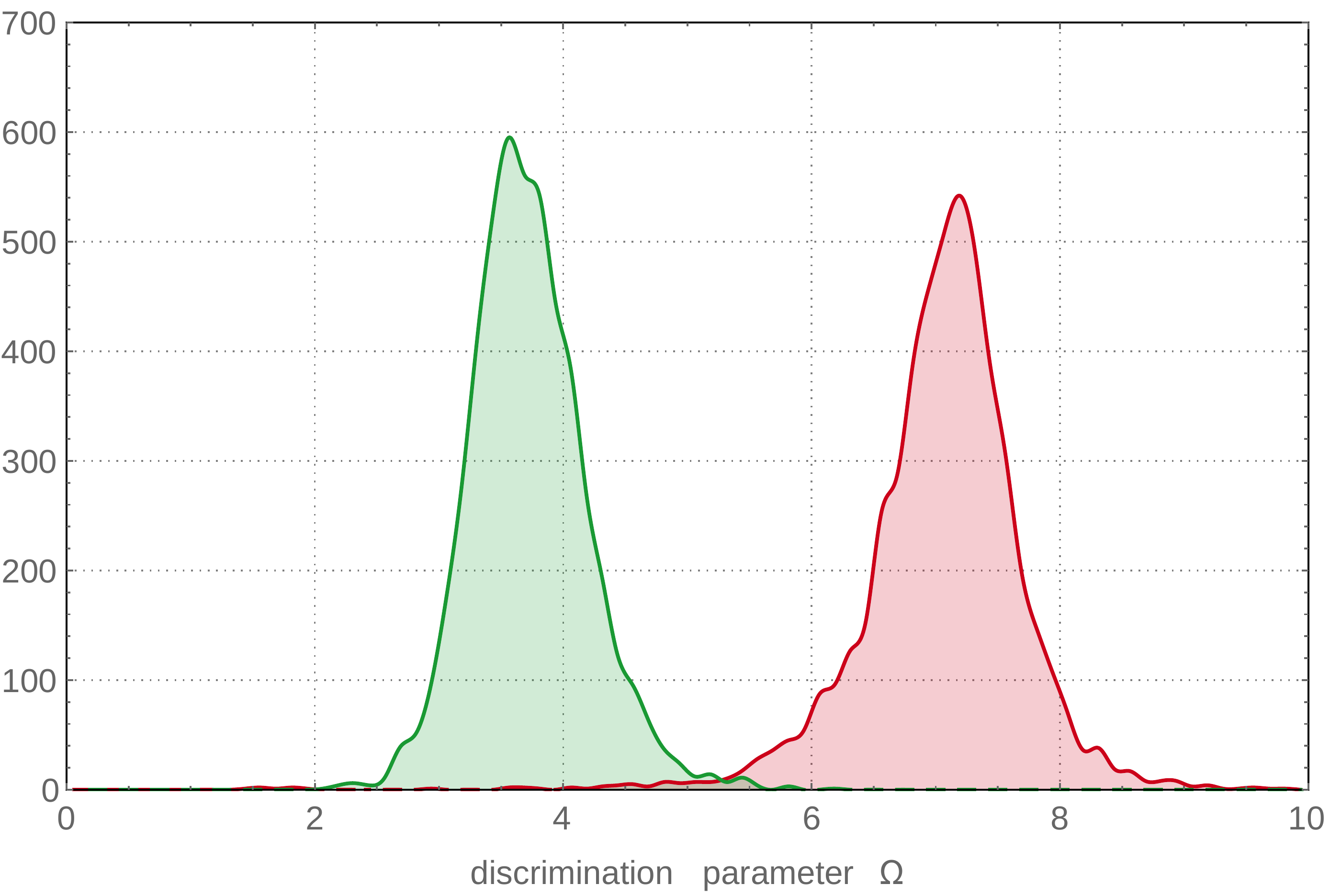}
		\hfill
		\includegraphics[width=0.49\textwidth]{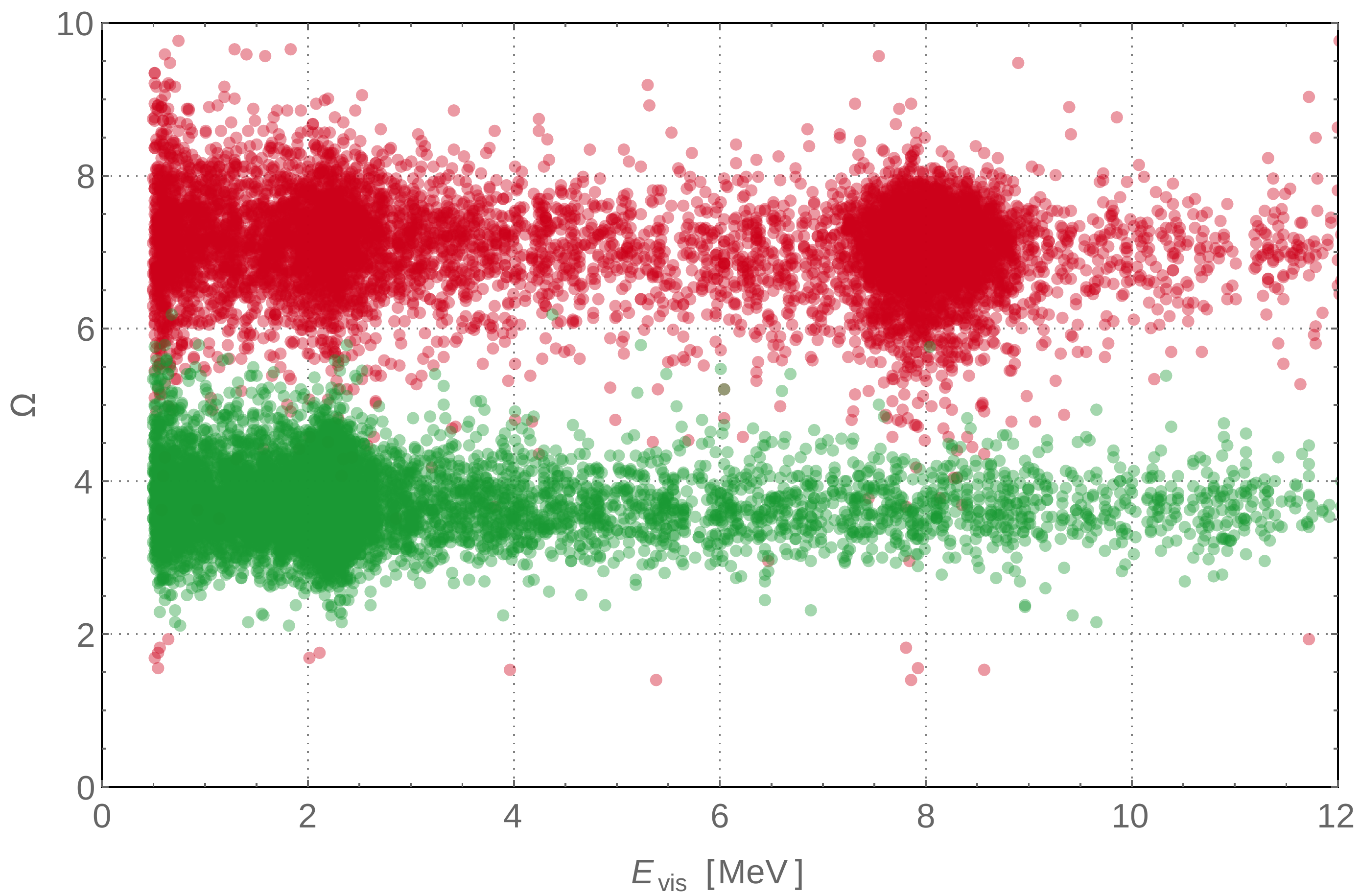}
		\caption{
		\emph{(Left)} Separation of Target (red) and Gamma Catcher (green) events with an optimized spectrum-based
		discriminator according to equation~\eqref{eq:OmegaOpt}. The data points were obtained from $^{252}$Cf calibration data
		in the detector center and in the middle of the Gamma Catcher at \mbox{$r=1429$~mm}.\label{img:GCNT}
		\emph{(Right)} Distribution of the discriminator values versus the visible energy. The larger populations at around 2.2~MeV are from deexcitation gammas after neutron capture on Hydrogen, which occurs in both volumes. In contrast, neutron capture on Gadolinium can only happen in the Target and the corresponding population at around 8~MeV only appears in the band belonging to Target events.\label{img:bands}}
		\end{figure}

The separation of GC and Target events serves as a demonstration of the capability of SSD to distinguish event categories exclusively with pulse shape information, as well as to illustrate that the parameter is independent of the visible energy and the event vertex.\\
\\
For this evaluation we used data from a $^{252}$Cf calibration source at the center of the Target volume and in the GC at a radial distance of $r= 1433.8$~cm from the detector center. $^{252}$Cf can undergo spontaneous fission and typically emits 2--4 neutrons with an average energy of around 2~MeV in the process.\footnote{In about 97\% of the cases $^{252}$Cf undergoes $\alpha$-decay, but the $\alpha$-particles do not penetrate the source encapsulation.} The neutrons can be specifically selected via their coincidence with accompanying gammas. In addition, the radioactive decay of the fission fragments provides uncorrelated events with up to around 15~MeV energy. Since we are interested in covering a wide range of energies, these events are desired in this study and no coincidence cut was applied, so this sample contains both neutrons and decay events. All events with visible energy $E_\mathrm{vis}\in[0.7, 12]$~MeV and a reconstructed vertex closer than 50~cm to the source position were selected. Events with a coincident signal in the Inner Veto of $Q_\mathrm{IV}>10000$~units of charge were tagged as muons and removed from the sample, as well as all events within 1~{\u}s after a muon. To remove light noise from the sample the standard cuts from Ref.~\cite{LN} were applied:
		(i)~\mbox{$q_\mathrm{max}/q_\mathrm{tot}<0.09$}, where $q_\mathrm{max}$ is the highest charge seen by a
		single PMT and $q_\mathrm{tot}$ is the total charge,
		(ii)~\mbox{$Q_\mathrm{dev}<30000$} units of charge, where
		$Q_\mathrm{dev}\coloneq \frac{1}{n} \sum_i^n \frac{(q_\mathrm{max}-q_i)^2}{q_i}$ and
		$i$ runs over all $n$ PMTs within a distance of 1~m from the PMT with the maximum charge $q_\mathrm{max}$, and
		(iii)~$\sigma_t<40$~ns, where $\sigma_t$ is the standard deviation of the PMT hit time distribution.

Figure~\ref{img:GCNT} shows the distribution of the optimized classifier for 5000 events in the Target (red) and in the Gamma Catcher (green) volume. It is seen in the left panel that the separation is nearly perfect, with an optimal cut position at $\Omega\approx5.2$. The right panel displays the classifier as a function of the visible energy. The values are distributed along horizontal bands, which indicates that $\Omega$ is to first order independent of the event energy. This is a result of the normalization of the scintillation pulse as described in Section~\ref{approach}. The confirmation of energy independence is important, since it reduces systematic uncertainties associated with cuts on $\Omega$. In particular, it is an essential requirement for use in $\theta_{13}$ analyses, since Double Chooz also utilizes the antineutrino energy spectrum to obtain $\theta_{13}$. A classifier with an energy dependence could potentially distort the spectrum and cause an additional systematic uncertainty.

		\begin{figure}[t]
		\centering
		\includegraphics[width=0.48\textwidth]{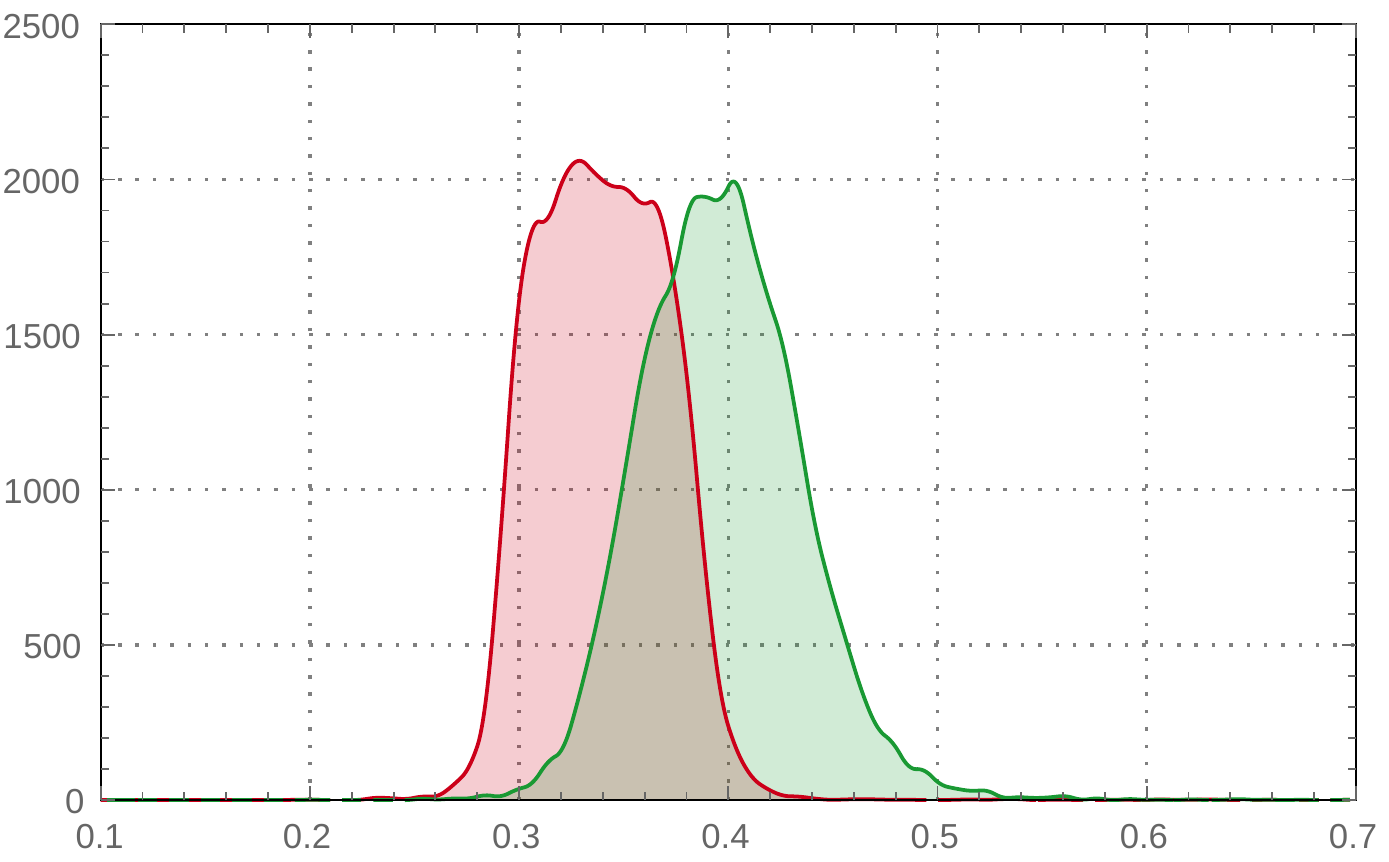}
		\hfill
		\includegraphics[width=0.481\textwidth]{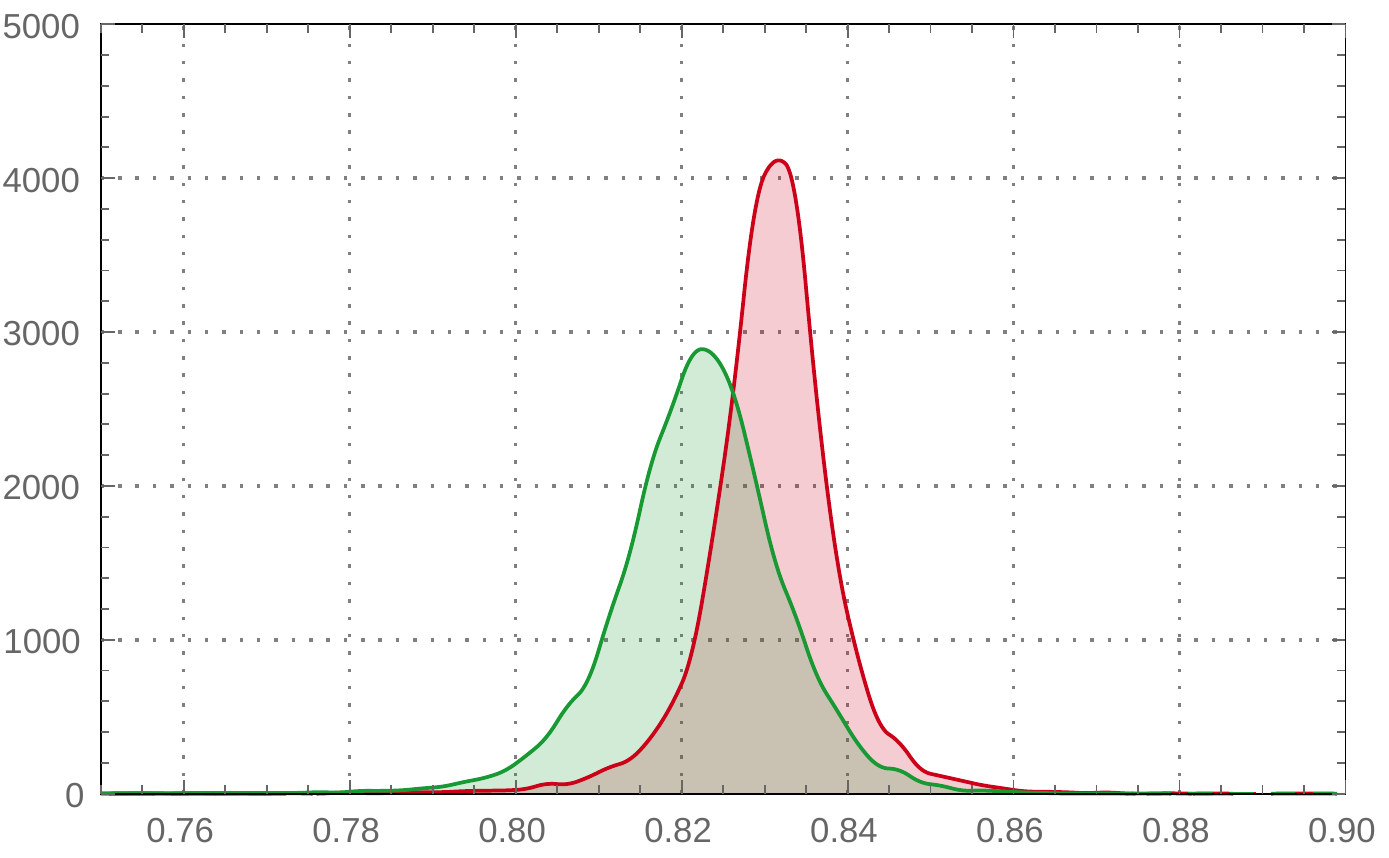}
		\\
		\includegraphics[width=0.482\textwidth]{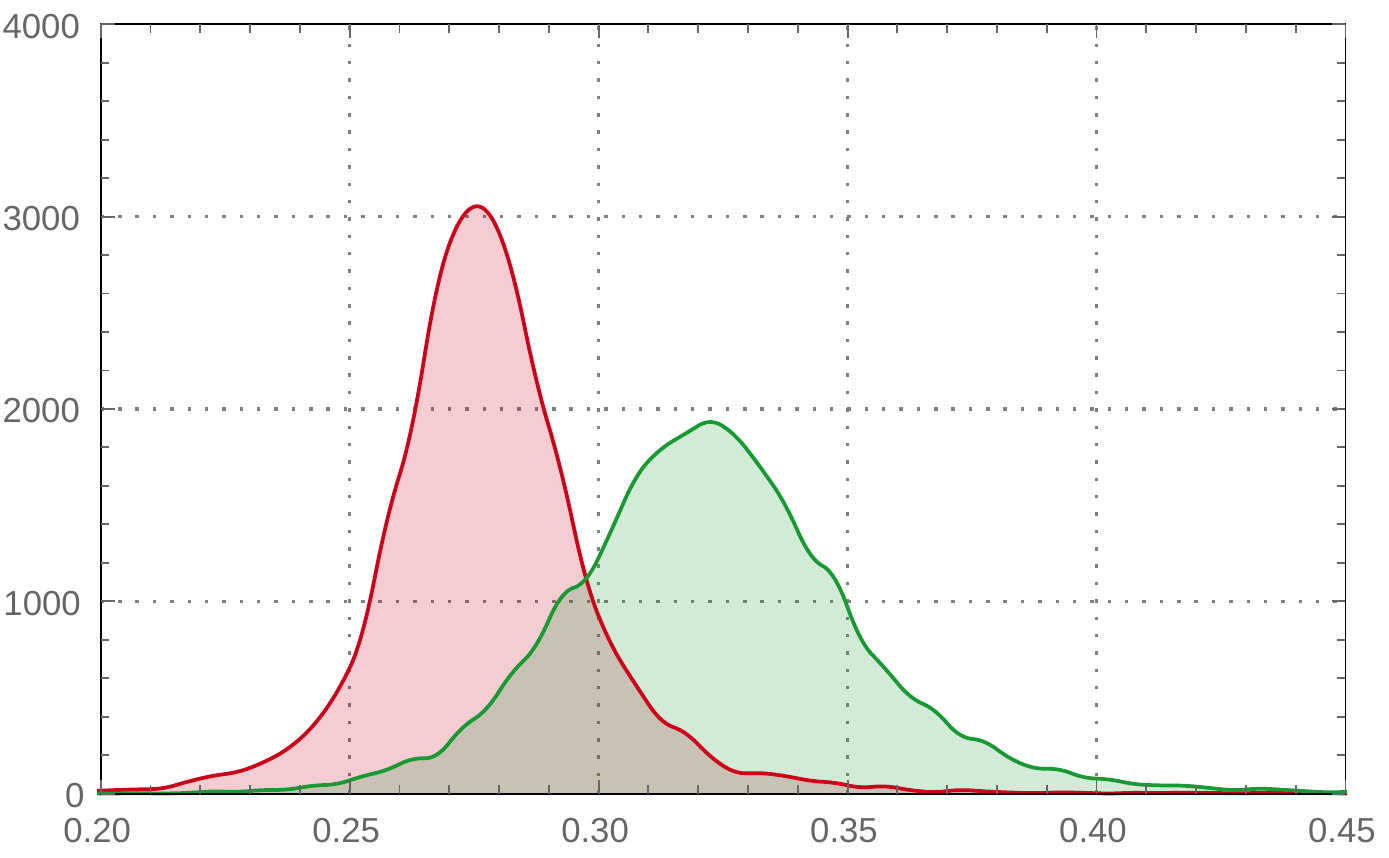}
		\hfill
		\includegraphics[width=0.466\textwidth]{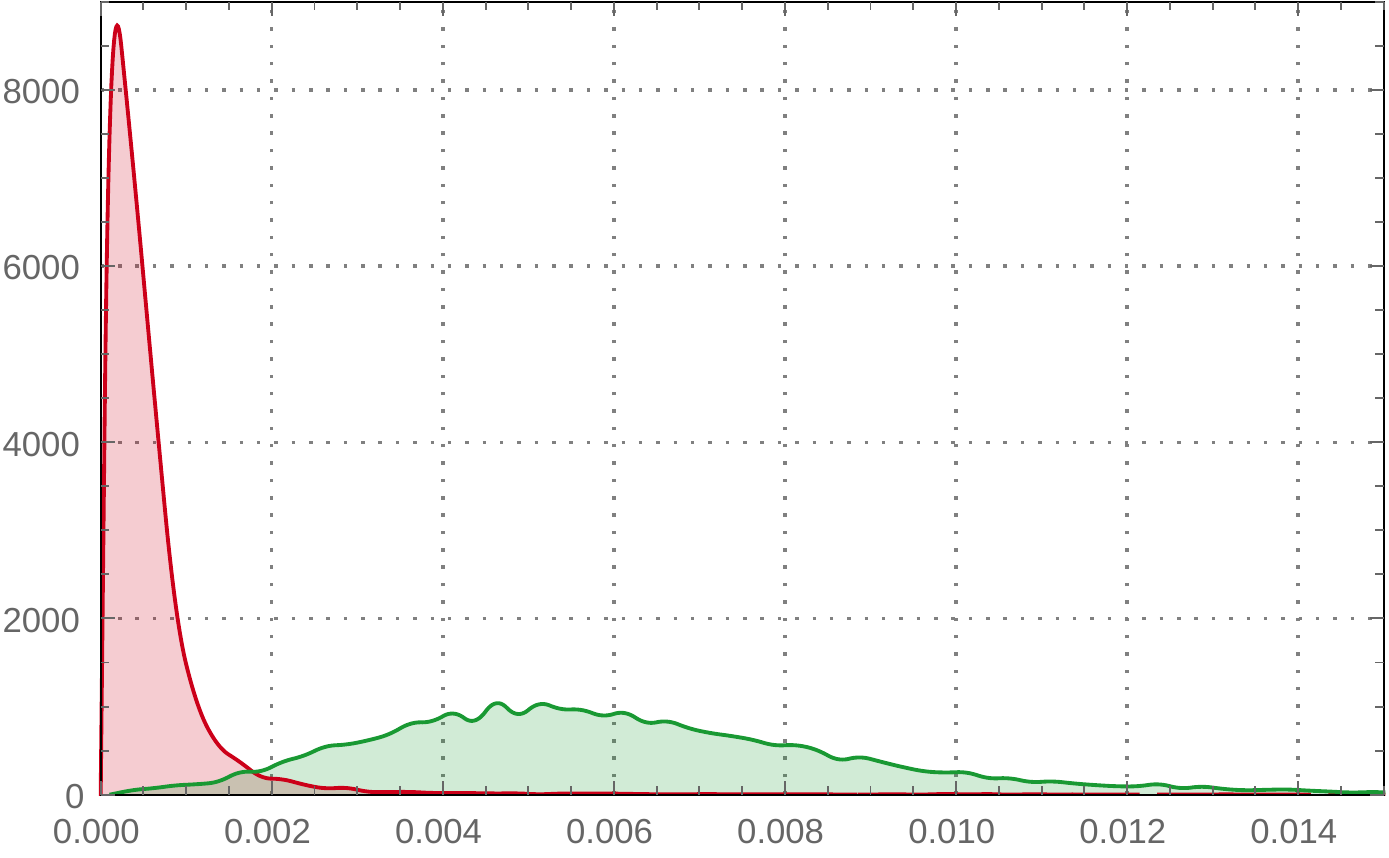}
		\includegraphics[width=0.8mm]{new_img/pixel.png}
		\caption{
		Distributions of commonly used PSD variables in the time domain for Target events (red) and GC events (green).
		Their corresponding figures of merit are given in Table~\ref{tab:figureofmerit}.
		For comparison, the distribution of the SSD classifier was shown in Figure~\ref{img:GCNT}.
		\emph{(Top left)}~comparison of the charge in the rising flank to the total charge,
		\emph{(Top right)}~comparison of the charge in the tail to the total charge,
		\emph{(Bottom left)}~comparison of the pulse amplitude 20~ns after the peak to the peak amplitude, and
		\emph{(Bottom right)}~similarity to a reference Target scintillator pulse shape.\label{classifierplots}
		}
		\end{figure}

The large population seen in the Target band at $\approx8$~MeV corresponds to neutron captures on Gadolinium. Since Gadolinium is only present in the Target but not the GC volume, this population is not present in the Gamma Catcher band. The bulge at $\approx2.2$~MeV corresponds to neutron captures on Hydrogen, which can occur in both volumes and are indeed seen in both the Target and the GC bands. Below 2.2~MeV both bands begin to spread in width, which is due to the smaller quantity of photons per event at lower energies, causing a larger statistical variation of the pulse shape, which is reflected in a wider range of $\Omega$ values.\\
\\
To better quantify the separation of Target (NT) and GC events we define a figure of merit
 $Q$ as
	\begin{equation}
	Q \coloneq \frac{\left| \mu_\mathrm{GC} - \mu_\mathrm{NT} \right|}{\sigma_\mathrm{GC} + \sigma_\mathrm{NT}},
	\label{Qdef}
	\end{equation}
where $\mu$ and $\sigma$ are the observed means and standard deviations of the classifier for GC and Target events.\footnote{The definition of the figure of merit $Q$ is somewhat arbitrary, but it should be a good measure for the discrimination performance of the classifier. Please note in particular that eq.~\eqref{Qdef} is not an operation on two random variables, but a description of the separation of their distributions. As such, the standard deviations in the denominator of eq.~\eqref{Qdef} do not necessarily have to be added quadratically. In fact, a quadratic summation of the standard deviations would add a bias and disfavor classifiers which yield large values, even if they provide a better separation. A linear addition instead guarantees that $Q$ remains unchanged if the classifier is scaled by a constant factor.} This definition compares the separation of the two distributions to their widths, given by the respective standard deviations. $Q$ is 1 when the error bars of the distributions just touch and becomes larger with increasing spacing between the error bars. If there is an overlap of the error bars, $Q$ is smaller than 1. The $\Omega$ discriminator achieved a figure of merit of $Q\approx3.56$ for the separation of Target and GC events.

The performance is now compared to several established PSD methods in the time domain, which were also tested with the same $^{252}$Cf calibration data. The $T_0$- and time of flight-corrected sum pulses (constructed according to steps 1--3 in Section~\ref{approach}) were smoothed with a Gaussian kernel in order to reduce the influence of noise and to improve the performance of the time-domain methods. Several commonly used approaches were tested and the results are summarized in Table~\ref{tab:figureofmerit}. The respective distributions of these classifiers are displayed in Figure~\ref{classifierplots}.

A popular method is \textit{charge comparison}, in which the charge contained in a certain time window over the pulse is compared to the total charge. With this method the best performance of $Q\approx 0.47$ was reached when the window extends from 20~ns after the peak to the end of the pulse. It is also possible to extend the window over the rising flank of the pulse, in which case $Q\approx0.44$ was obtained. \textit{Pulse gradient analysis} compares the pulse amplitude at a specific time to the peak amplitude. This method yielded $Q\approx1.16$ when the test point was chosen at 20~ns after the peak.

The best performance of the time-domain techniques was achieved by comparing each pulse to a reference pulse shape. A reference pulse for the Target scintillator was obtained by averaging over all valid pulse shapes from a calibration source in the detector center. A standard $\chi^2$ approach was then used to quantify the similarity of individual pulses to the reference pulse, which yielded $Q\approx1.42$.

		\begin{table}[t]
		\centering
		\caption{\label{tab:figureofmerit} Figure of merit $Q$ for the separation of Target and GC pulses for different PSD methods in the time domain, as well as for the SSD technique.}
		\smallskip
		\begin{tabular}{|l l|}
		\hline
		Method						& $Q$\\
		\hline
		Charge Comparison (flank)		& 0.44\\
		Charge Comparison (tail)		& 0.47\\
		Pulse Gradient Analysis			& 1.16\\
		Comparison with reference		& 1.42\\
		SSD							& 3.56\\
		\hline
		\end{tabular}
		\end{table}

\begin{figure}[t]
		\centering
		\includegraphics[width=0.7\textwidth]{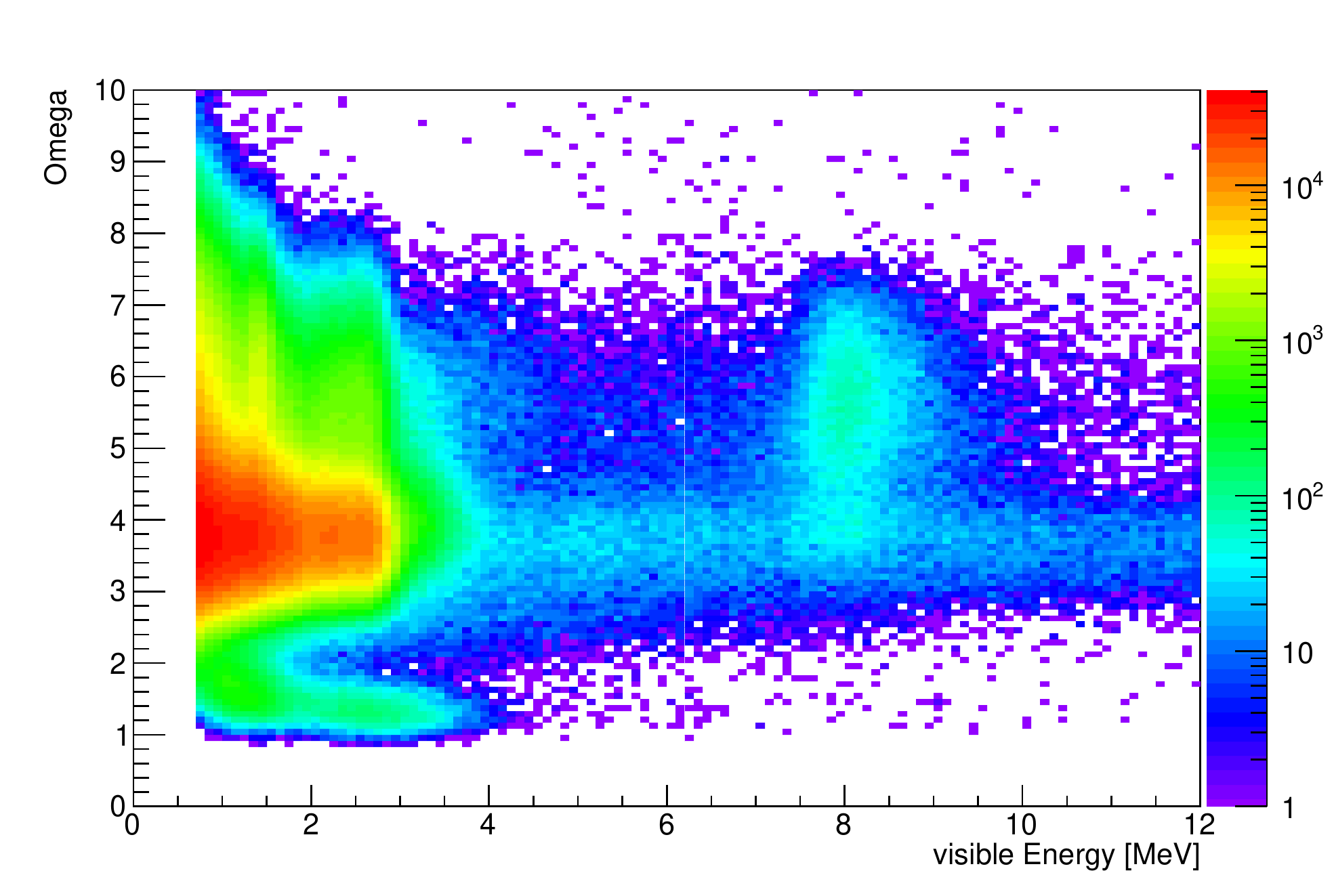}
		\caption{
		Distribution of the discriminator values versus the visible energy $E_\mathrm{vis}$ for singles events after the
		preselection. Like for calibration data in Figure~\ref{img:bands}, the different bands are visible. Different cuts on the
		discriminator variable are able to select different event categories from this distribution (see Figures~\ref{fig3GC}
		and~\ref{fig3NT}).\label{fig3}}
		\end{figure}

Even though the difference between GC and Target pulse shapes are relatively large, the tested time-domain methods did not achieve a separation comparable to the SSD technique. However, it is worth noting that the figure of merit only indicates the performance of a specific method under the given experimental conditions. In other environments the results may vary.\\
\\
The SSD technique is now applied to physics data and used to distinguish different classes of events with cuts on the discriminator variable. For this purpose a sample of \textit{singles} events was studied, i.e. a sample of all valid physics events before applying any coincidence criteria. The selection criteria were the same as for the calibration data before, but included the whole detector volume (no cuts on the event position were applied). Figure~\ref{fig3} shows the distribution of the classifier against the visible energy of singles events. In contrast to calibration data in Figure~\ref{img:GCNT} the Target band has mostly vanished, because singles events are dominated by natural radioactivity, which rarely have energies above 3~MeV. The Gamma Catcher still shows a faint horizontal line up to higher energies, which is caused by fast neutrons that are created by muon-induced spallation reactions in the rocks surrounding the detector. These neutrons may enter the Gamma Catcher and create visible proton recoils in a wide range of energies.

Thermal neutrons can be captured by Gadolinium in the Target volume. The corresponding cluster at $E_\mathrm{vis}\approx8$~MeV is still visible in the Target volume. After capturing a neutron Gadolinium deexcites via emission of four gammas with $\approx2$~MeV each. If a capture event occurs close to the Target boundary, some of the gammas can enter the Gamma Catcher and deposit their energy there. In this case the resulting scintillation pulse shape is a superposition of the Target and Gamma Catcher pulse shapes, which is why in Figure~\ref{fig3} the population appears to ``leak'' into the Gamma Catcher band.

\begin{figure}[tp]
		\centering
		\includegraphics[width=0.49\textwidth]{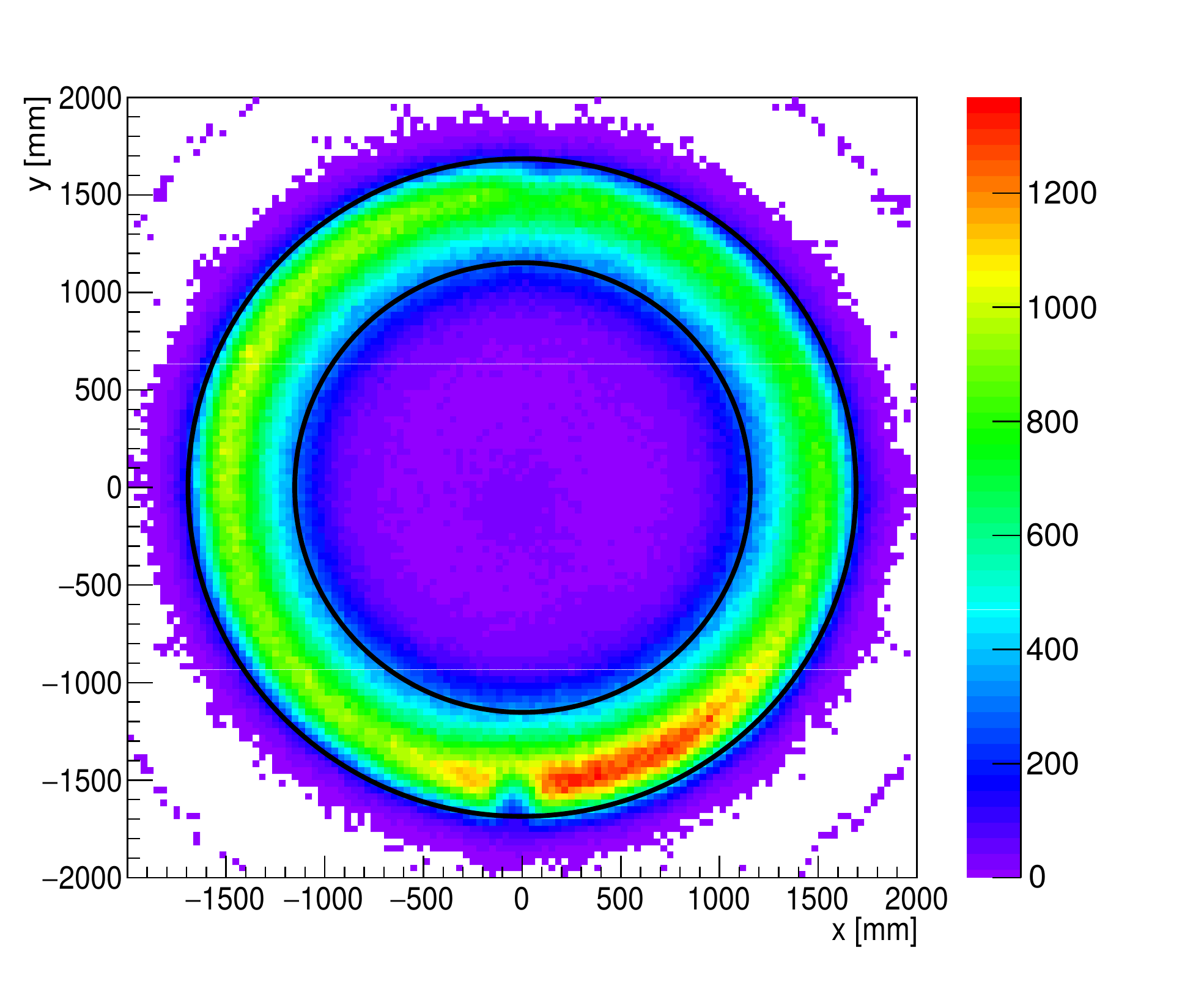}
		\hfill
		\includegraphics[width=0.49\textwidth]{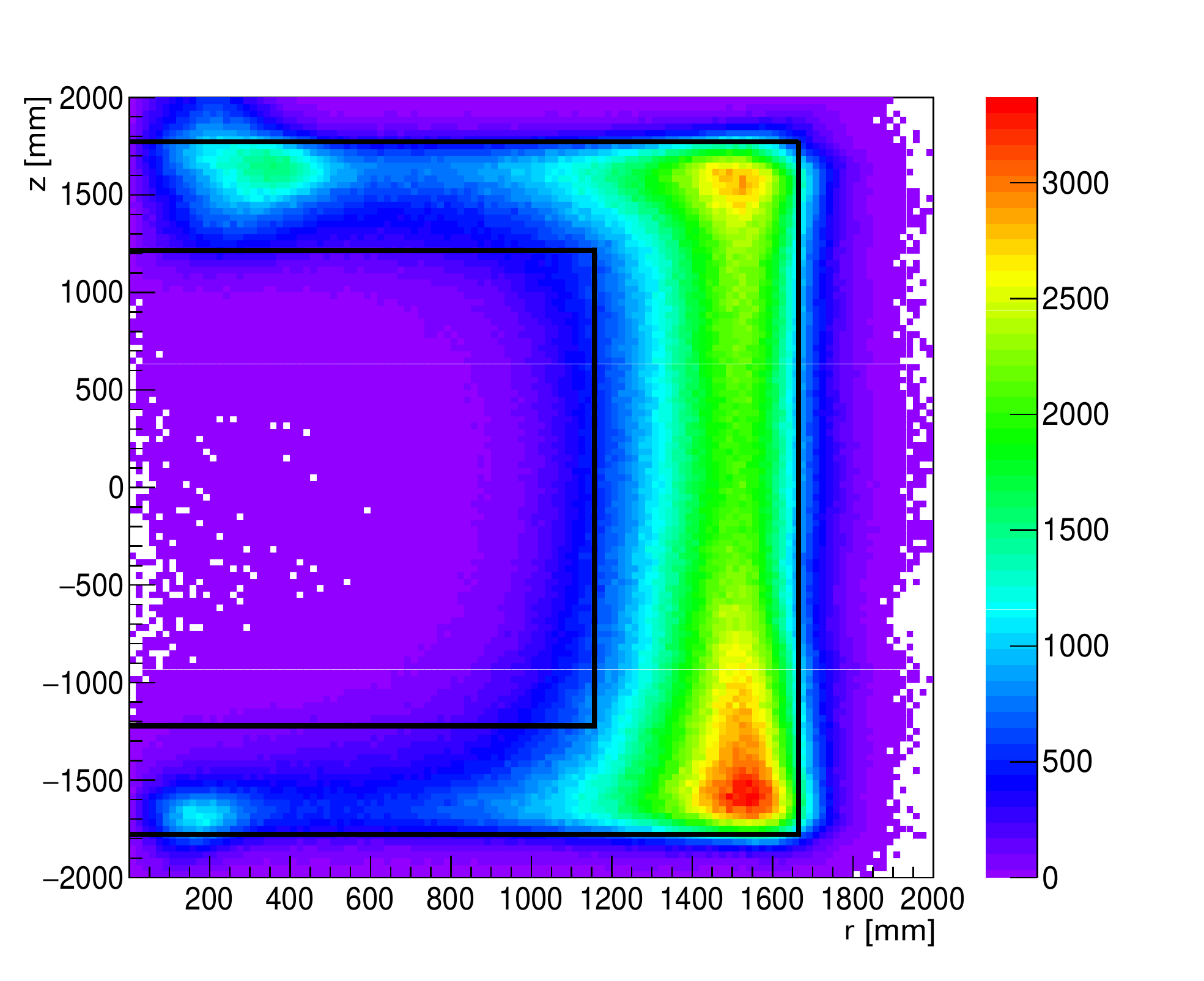}
		\caption{
		Vertices of singles events with $E_\mathrm{vis}\in[0.7, 12]$~MeV after applying the cut $2.0 < \Omega < 5.2$.
		The black lines represent the boundaries of the Target and the Gamma Catcher volumes.
		The cut on $\Omega$ selects predominantly Gamma Catcher events.
		\emph{(Left)} Top-down view of the selected event vertices. For illustration purposes, an additional cut
		$|z| < 1000$~mm has been applied, so that the GC lids do not appear in this image.
		\emph{(Right)} Two-dimensional vertical vs. radial distribution of the reconstructed vertices.\label{fig3GC}}
		\includegraphics[width=0.49\textwidth]{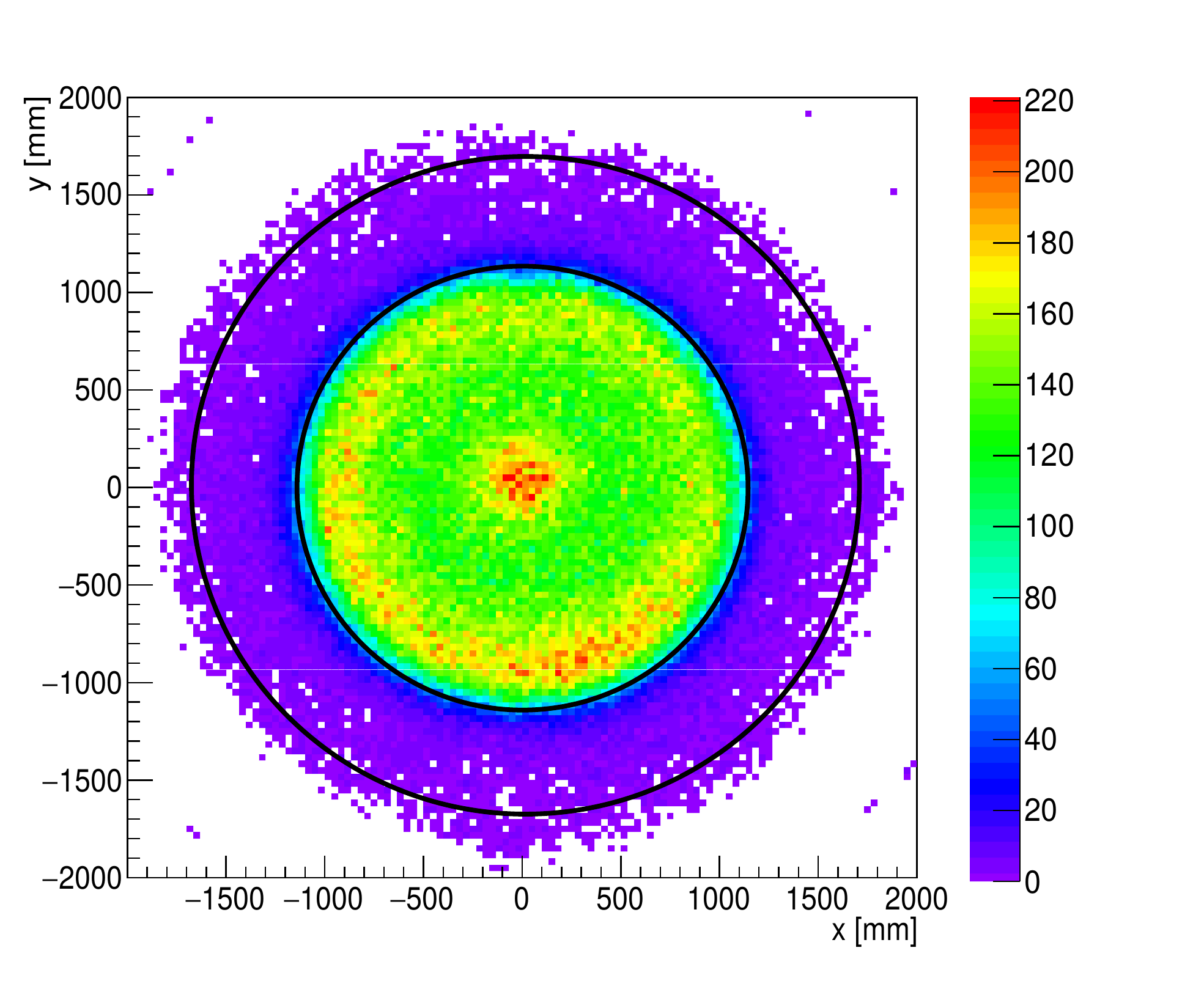}
		\hfill
		\includegraphics[width=0.49\textwidth]{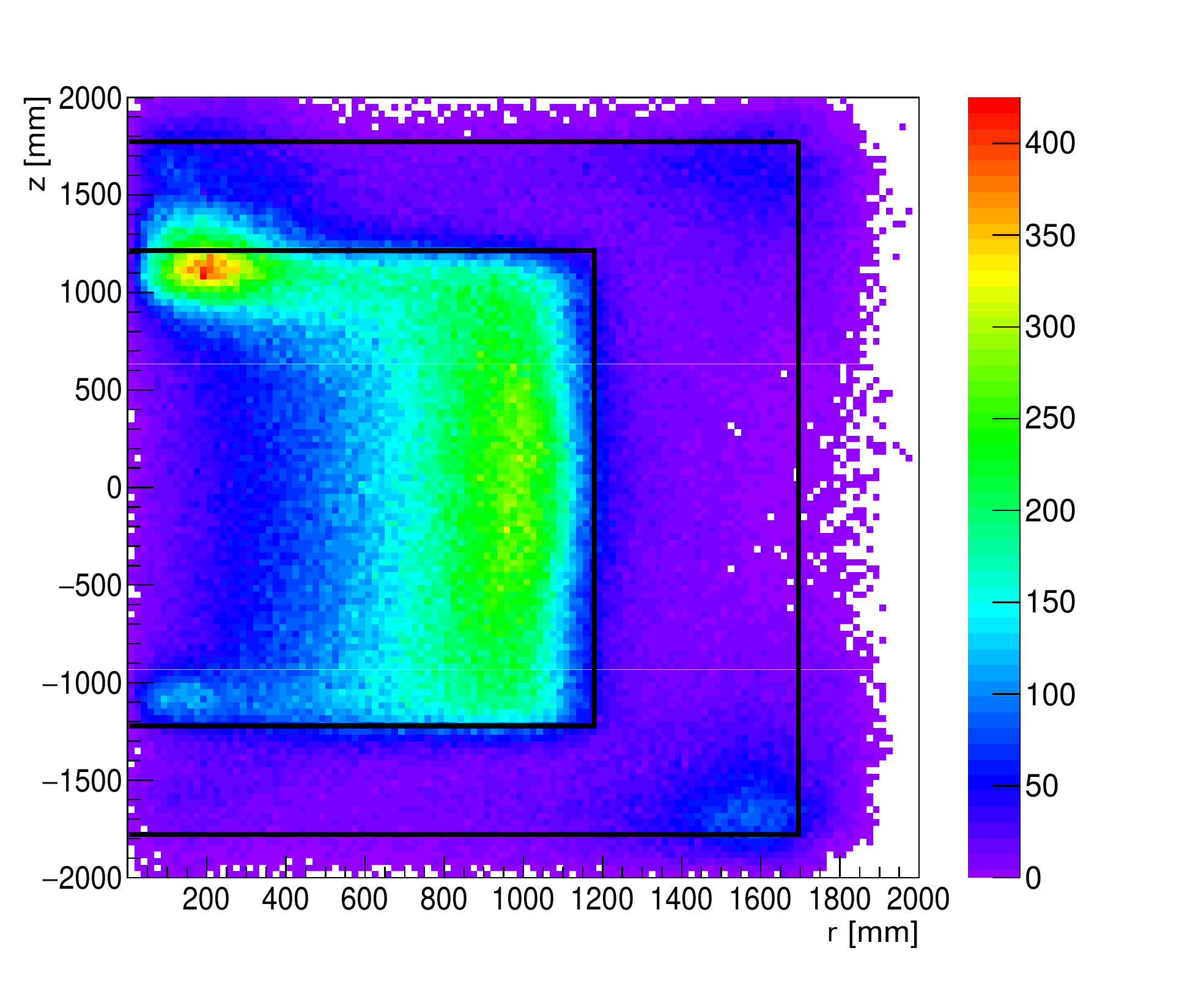}
		\caption{
		Vertices of singles events with $E_\mathrm{vis}\in[0.7, 12]$~MeV after applying the cut $\Omega>5.2$.
		The black lines represent the boundaries of the Target and the Gamma Catcher volumes.
		The cut on $\Omega$ selects predominantly Target events.
		\emph{(Left)} Top-down view of the selected event vertices. For illustration purposes, an additional cut
		$|z| < 1000$~mm has been applied, so that the GC lids do not appear in this image.
		\emph{(Right)} Two-dimensional vertical vs. radial distribution of the reconstructed vertices.\label{fig3NT}}
		\end{figure}

At $\Omega<2$ another population is visible, which is due to light noise events that survived the selection cuts, as will be shown in Section~\ref{sec:ln}. The fact that they appear as a separate cluster in $\Omega$ provides a new handle to identify them.\\
\\
This distribution of singles events in Figure~\ref{fig3} can be separated into different event categories by applying cuts on the discriminator $\Omega$. Figure~\ref{fig3GC} shows the vertex position of the events selected with $2.0 < \Omega < 5.2$, where the lower value was applied to remove the light noise population. The reconstructed event vertices lie almost exclusively in the GC volume.

In contrast, the cut $\Omega > 5.2$ selects predominantly Target events, as shown in Figure~\ref{fig3NT}. Since singles events are mainly caused by externally incident radioactivity, they accumulate at the outer regions of the Target volume. The very sharp division between the Gamma Catcher and the Target volumes is proof that $\Omega$ is indeed sensitive to the different scintillation pulse shapes in both volumes, and that its separation capability is not merely an effect of the position reconstruction or the time of flight correction.

By selecting only singles events within different confined regions (e.g. at the top and the bottom of the volumes) it was also confirmed that the classifier does not depend on the position of the event vertex. This is expected, since the primary pulse shapes produced by the scintillator are equal throughout the volume and the time of flight correction removes position-induced distortions. This also means that the classifier only provides binary information about the detector volume, but cannot give information about the exact event vertex. On the other hand, the vertex reconstruction algorithm used in Double Chooz has a resolution of the order of 10~cm~\cite{DC2}. If an event occurred very close to a boundary, it is unable to decide in which volume the event took place. In such cases $\Omega$ can add to the knowledge of the event position.\\
\\
This study has demonstrated that the SSD technique is able to separate event categories exclusively by their pulse shapes. Other methods routinely employed to categorize events in large-scale liquid scintillator detectors usually make use of charge and time information. It is shown that the pulse shape contains additional and independent information about events, which can be successfully exploited to gain additional information. This opens up new possibilities for data analysis, especially concerning particle identification (PID) with help of the scintillation pulse shape.

\subsection{Light noise rejection}\label{sec:ln}
		\begin{figure}[tp]
		\centering
		\includegraphics[width=0.49\textwidth]{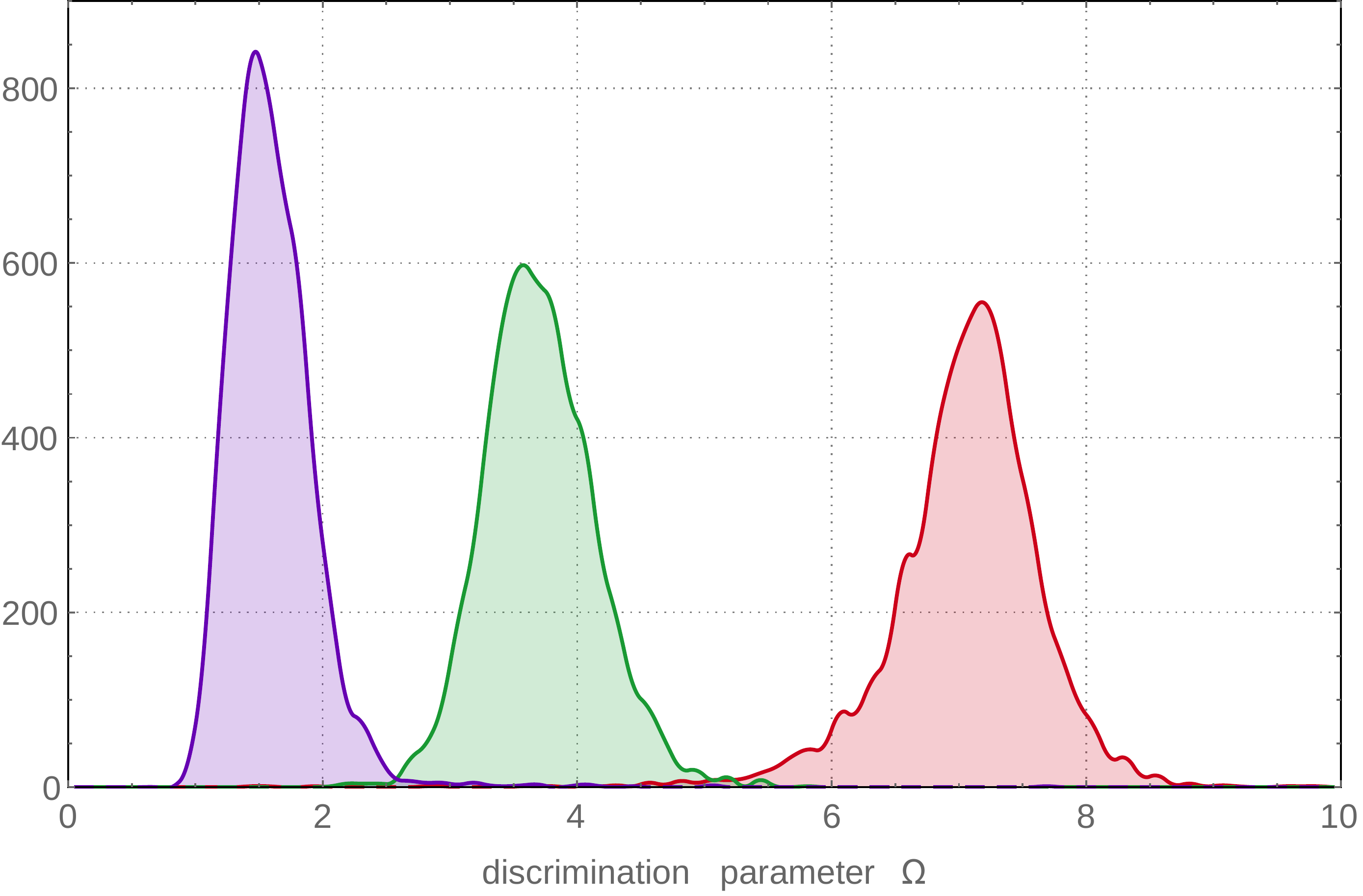}
		\hfill
		\includegraphics[width=0.49\textwidth]{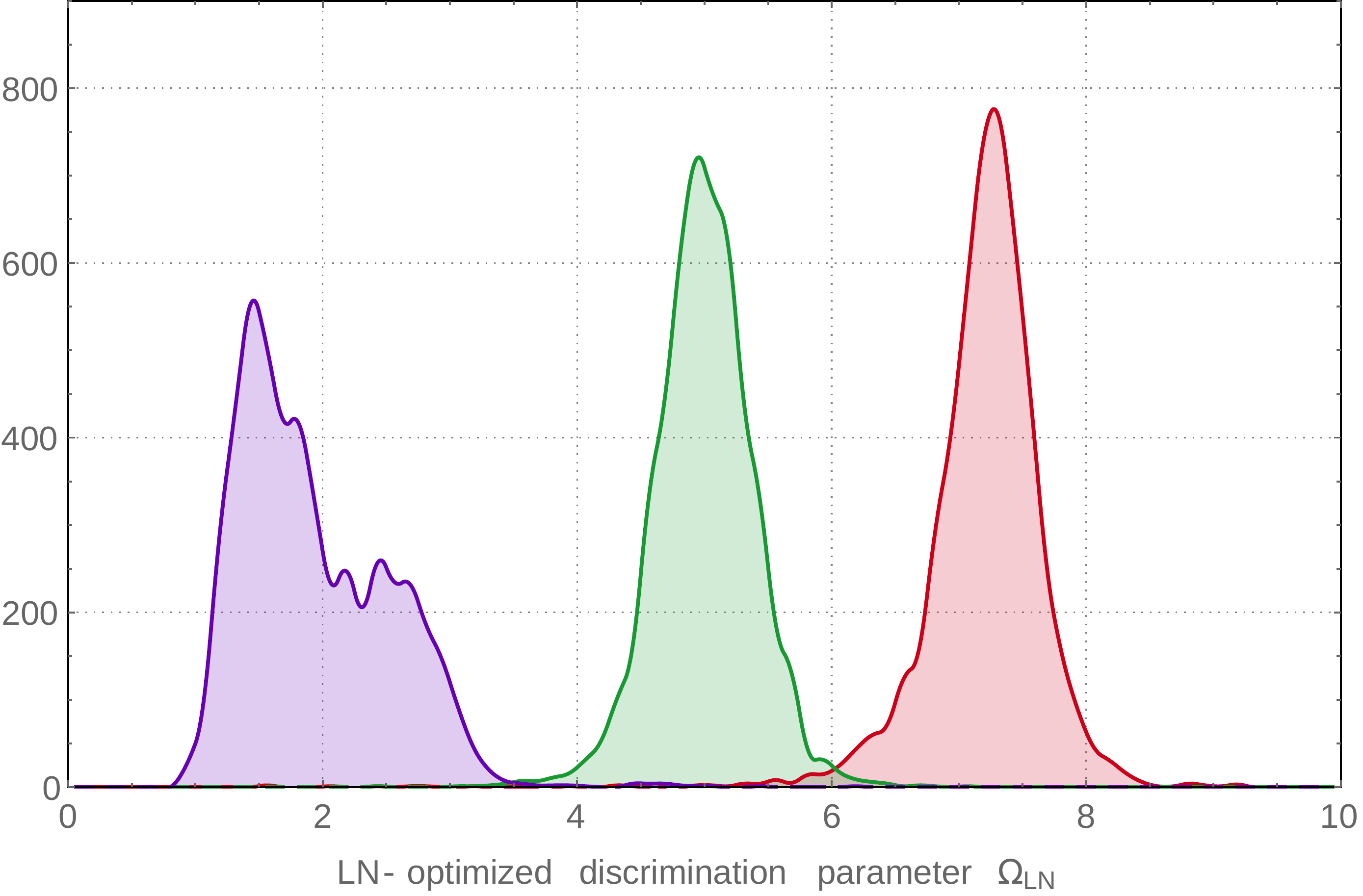}
		\caption{
		Distribution of different discriminator variables for light noise (blue), Gamma Catcher (green) and Target (red) events.
		\emph{(Left)} This panel shows the distribution of the parameter $\Omega$, which is optimized for GC/Target separation
		and was used in Section~\ref{sec:GCTargetSeparation}.
		\emph{(Right)} Distribution of a discriminator $\Omega_\mathrm{LN}$ specifically optimized for light noise identification.				\label{LNoptimization}}
		\end{figure}
\textit{Light noise} is a type of instrumental background where electric discharges in a PMT base produce light, which can enter the detector and be seen by other PMTs, causing a non-physical background in the detector data. Such events are usually characterized by a very irregular signal shape and have a very narrow power spectrum. The phenomenon and its role in Double Chooz are described in detail in Ref.~\cite{LN}.\\
\\
Due to their vastly different pulse shapes in comparison to physics events, the SSD technique can be used very efficiently to identify light noise events. For this investigation, a sample of light noise events was selected with the dedicated cuts described in Ref.~\cite{DC3}:
		(i)~$q_\mathrm{max}/q_\mathrm{tot} > 0.12$,
		(ii)~$Q_\mathrm{dev} > 30000$ units of charge, and
		(iii)~$\sigma_t > 36\,$ns and $\sigma_q > (464-8\,\sigma_t)$ units of charge, where $\sigma_t$ and $\sigma_q$ are the
		standard deviations of the hit time and charge distributions, respectively.
Figure~\ref{LNoptimization} shows the distribution of the standard classifier $\Omega$ optimized for GC/Target separation, which was used in Section~\ref{sec:GCTargetSeparation}, and compares it with a classifier $\Omega_\mathrm{LN}$ specifically optimized for the separation of light noise from physics events. The Fourier power spectra of light noise and physics events show a best separation at different coefficients than the spectra of GC and Target events. The $\Omega_\mathrm{LN}$ classifier contains the optimal weights for this task. It is also seen that the light noise curve is changing its shape after optimization, which is another indication that light noise is a collection of several different processes, rather than a single, well-defined phenomenon. Changing the weights of the Fourier coefficients affects these individual processes differently, leading to the apparent distortion of the curve.

		\begin{figure}[t]
		\centering
		\includegraphics[width=0.5\textwidth]{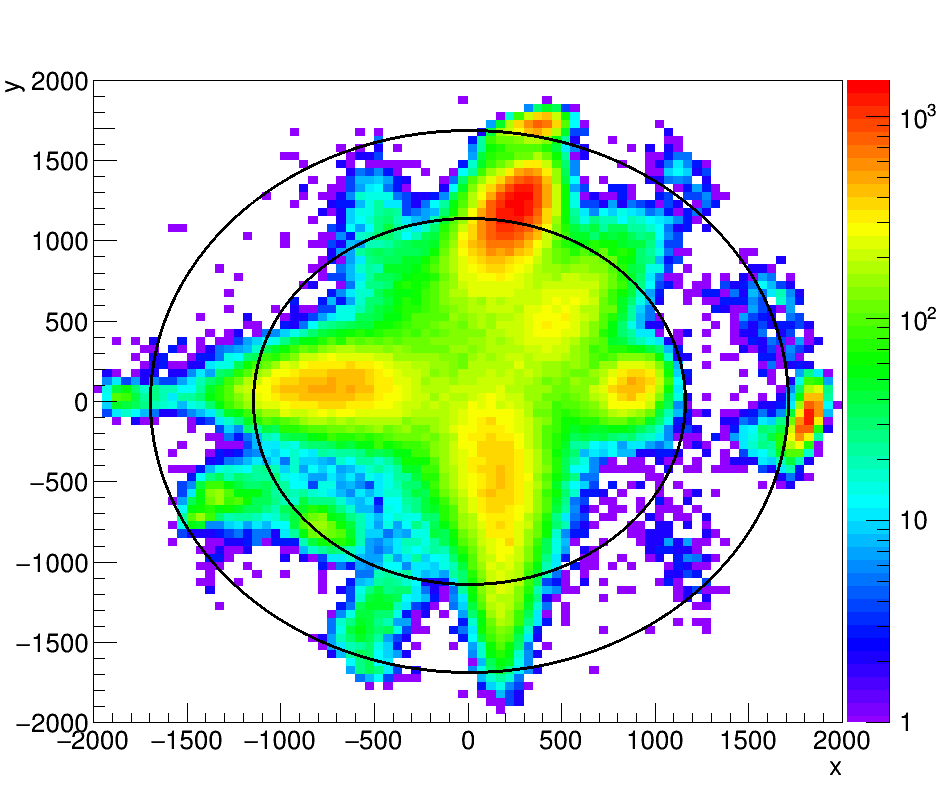}
		\caption{Vertices of singles events after a cut $\Omega_\mathrm{LN} < 2$. The black lines represent the volume
		boundaries of the Target and the Gamma Catcher.\label{fig3LN}}
		\end{figure}
		
While the standard discriminator $\Omega$ already separates light noise from physics events, the performance is improved further by $\Omega_\mathrm{LN}$. The discriminator can therefore serve to remove a large fraction of the remaining light noise without significantly affecting physics events. The current pre-selection cuts also remove most of the light noise populations, with the notable exception of a cluster, which is still present in the singles events and seen in Figure~\ref{fig3}. This specific population is untouched by any of the LN variables currently in use, resulting in evidence that there is a remaining light noise contamination in singles data that can be removed with the SSD technique.

Since there is essentially no physics with $\Omega_\mathrm{LN} < 3.5$, this cut allows a very clean rejection of LN events. The events rejected by this cut exhibit clear characteristics of light noise. Figure~\ref{fig3LN} shows the reconstructed event vertices of these events. The irregular vertex distribution of these events is a strong indications of light noise. In addition, most of these populations vanish after applying established light noise cuts. Their appearance is often intermittent (i.e. they show up in some runs but not in others), or they vanish after a specific PMT is switched off (as was the case with very LN-prone PMTs), accumulating to strong evidence that these populations are in fact light noise events.\\
\\
In studies of $\theta_{13}$ a spectral-based light noise rejection can help to reduce the number of accidentals even below current levels. Due to the high delayed energy and the short coincidence time the Gd-selection is already very clean and the remaining light noise events, as selected with $\Omega_\mathrm{LN}$, account only for about 0.6~\% of all events. However, due to the lower delayed event energy and larger coincidence window, there are more light noise events left in the Hydrogen selection. About 1.5~\% of the events can be identified as light noise candidates with $\Omega_\mathrm{LN}<3.5$, even after additional background rejection techniques were applied. In comparison, the estimated total amount of accidentals left in the Hydrogen selection is about 6~\%, so the SSD approach could remove about a quarter of the remaining accidentals~\cite{H2}. While the remaining light noise events do not have an impact on the result of a $\theta_{13}$ analysis, an integration of SSD may allow a relaxation of current selection criteria and help to improve the sensitivity.

\subsection{Particle identification}\label{sec:PID}
The characteristic energy deposition functions $\left<\mathrm{d}E/\mathrm{d}x\right>$ of different particle species cause the excitation of different amounts of singlet and triplet states in the scintillator, leading to differences in their pulse shapes, which can be exploited by several established techniques. The possibility of using pulse shape differences to infer the particle type is very attractive for liquid scintillator experiments, since it can help to identify and reject backgrounds.

However, PSD usually only works efficiently in detectors which were designed for that purpose, e.g. Borexino. Double Chooz, on the other hand, was not designed with a focus on strong PSD capabilities and the performance of established methods is limited. Nevertheless, as demonstrated in Section~\ref{sec:GCTargetSeparation}, the SSD-based classifier outperformed traditional methods in the task of separating GC, Target and light noise events and it is expected to yield some information about the particle type as well.

Since the pulse shape differences between different particles are much smaller than those between the GC and Target, the following studies were restricted to the Target volume by means of vertex cuts $R<1$~m and $|z|<1$~m.

\paragraph{Muons and stopping muons}
Muons show very characteristic values of the discriminator. They typically have long tracks and high energy depositions in the detector. Their pulses are generally broad and often also distorted by clipping, leading to very low $\Omega$ values in the range of light noise events ($\Omega$<2.5). They can be efficiently separated from other physics events with standard cuts and are not further investigated here.

This is not the case for \textit{stopping muons}, i.e. muons that stop within the detector and only deposit a limited amount of energy. Some muons can enter the detector undetected through the \textit{chimney}, a small tube on top of the DC detector that is used for filling and calibration (see Ref.~\cite{DC2}). If a muon enters the detector through the chimney (bypassing the Inner Veto) and stops in the uppermost part of the detector (so the visible energy deposition in the Inner Detector is not large enough for it to be identified as a muon) it can produce a valid prompt signal.

		\begin{figure}[t]
		\includegraphics[width=0.49\textwidth]{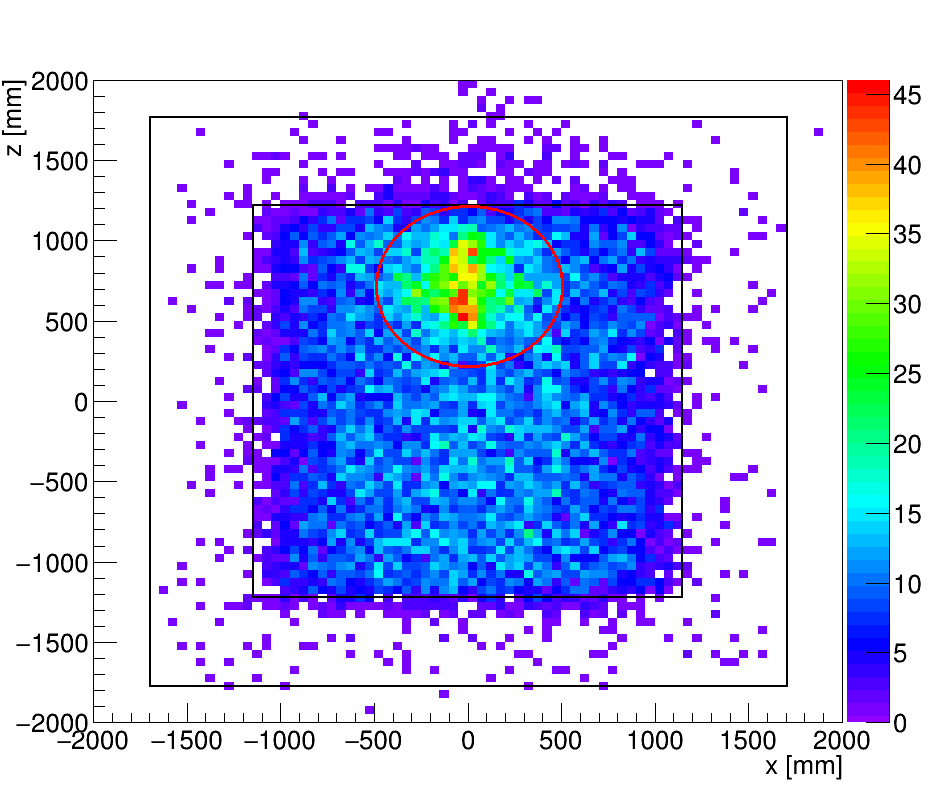}
		\hfill
		\includegraphics[width=0.49\textwidth]{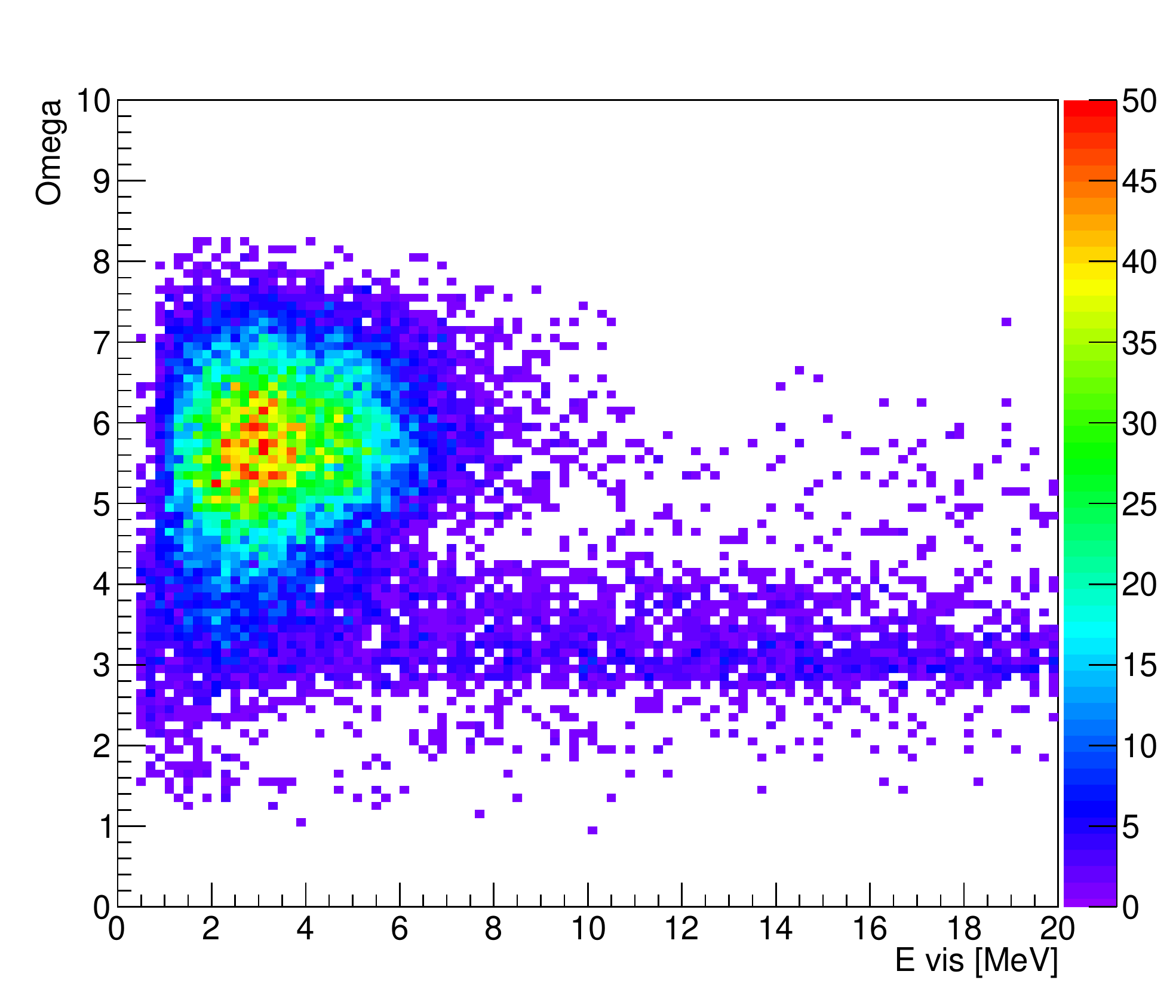}
		\caption{
		\emph{(Left)} Reconstructed vertices for the prompt events of a Gadolinium selection with relaxed cuts
		from page~\pageref{SMcuts} to artificially increase the number of stopping muon events. The sample shows an
		accumulation of events about halfway between the chimney and the detector center (circled in red).\label{img:SMvertices}
		\emph{(Right)} The discriminator values in dependence of the visible energy $E_\mathrm{vis}$ for the events 
		shown in the left panel.
		\label{4b}}
		\end{figure}

In muon decays at rest, the subsequent Michel electron provides a valid candidate for a delayed event if the visible energy of the event lies within the selection window. For this reason stopping muons are an important correlated background for IBD searches.

For technical reasons the chimney region cannot provide an efficient particle detection. When an event takes place in the chimney, part of the scintillation light is blinded by detector structures, so that mainly light emitted in a downwards cone can be seen. This screening leads to an incorrect vertex reconstruction as well as a significant distortion of the scintillation pulse shapes. While in principle undesired, this effect can be used to identify stopping muons and several analysis methods already make use of it~\cite{DC3}.\\
\\
For the investigation of stopping muons with the SSD technique a sample of IBD candidates was used, which were selected from the singles events based on the criteria defined in Ref.~\cite{DC3}, but with a modified coincidence time window to increase the stopping muon contribution:\label{SMcuts}
		(i)~the visible energy of the prompt event satisfies $E_\mathrm{vis}\in[0.5,20]$~MeV,
		(ii)~the visible energy of the delayed event satisfies $E_\mathrm{vis}\in[4,13]$~MeV,
		(iii)~a time coincidence between the prompt and delayed event of $\Delta T \in [0, 150]$~{\u}s,
		(iv)~a distance between the prompt and delayed reconstructed vertices of $\Delta R < 100$~cm, and
		(v)~no additional events are encountered within a time window of $[-200,600]$~{\u}s around the prompt signal.
The additional background reduction criteria defined in Ref.~\cite{DC3} were not applied in order to keep stopping muons in the sample. The resulting IBD candidate sample is thus expected to have a noticeable contamination of stopping muon events.
		\begin{figure}[t]
		\includegraphics[width=0.49\textwidth]{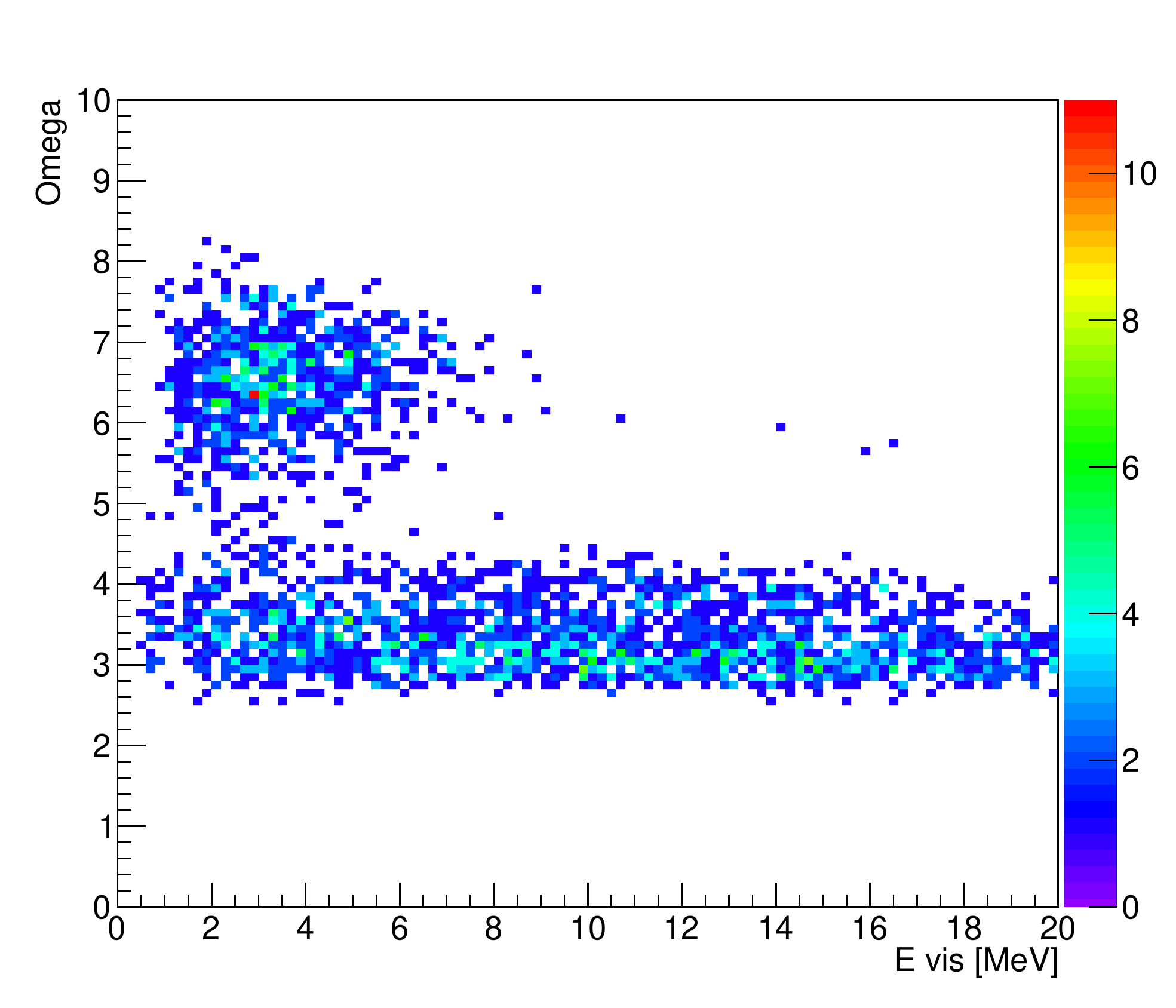}
		\hfill
		\includegraphics[width=0.49\textwidth]{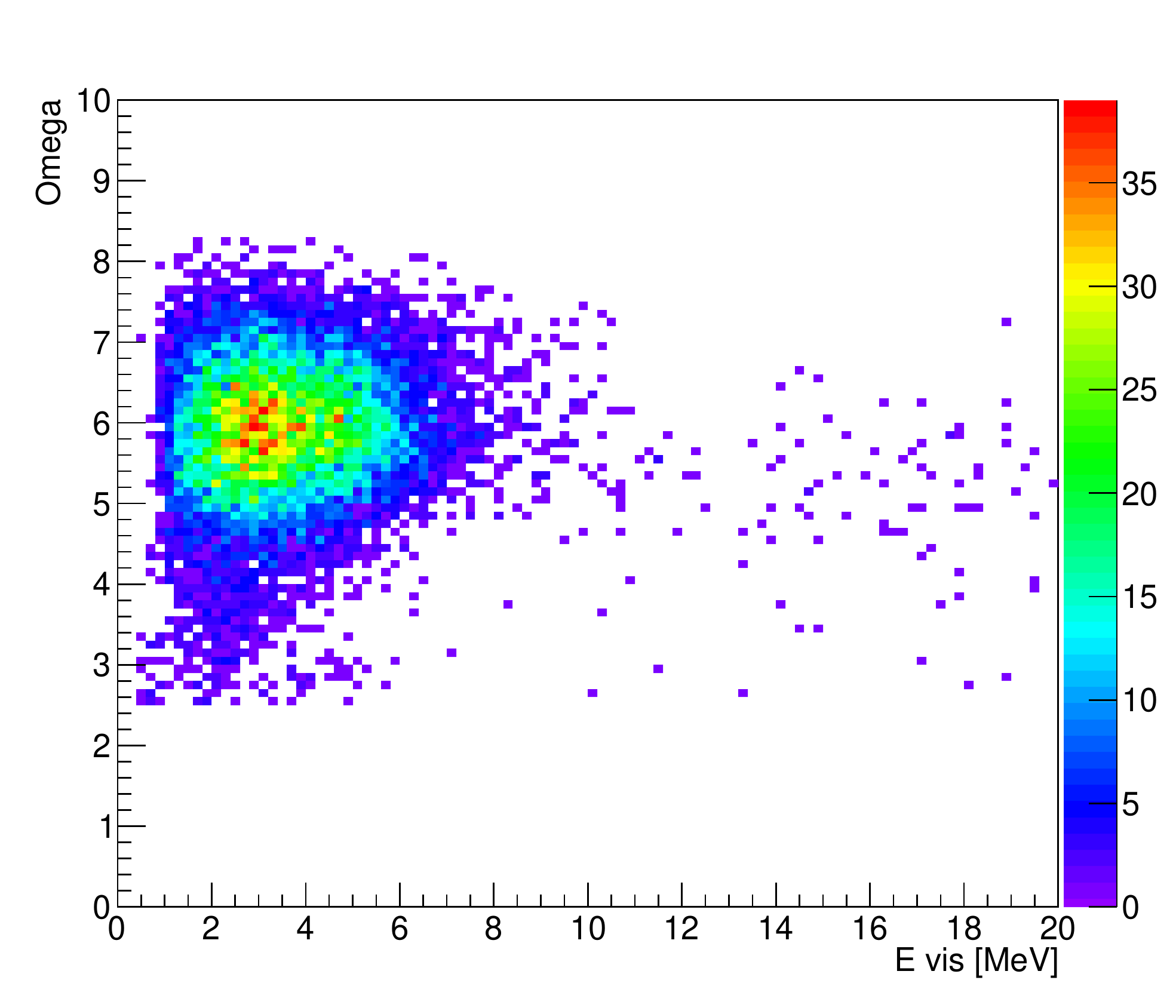}
		\caption{The discriminator variable $\Omega$ for \textit{prompt} events with a reconstructed vertex
		position inside and outside a sphere of radius $\Delta R<50$~cm around the point $z=75$~cm
		(shown as a red circle in Figure~\ref{4b}).
		\emph{(Left)} Distribution of $\Omega$ for prompt events inside the sphere.\label{4c}
		\emph{(Right)} Distribution of $\Omega$ for prompt events outside the sphere.\label{4d}}
		\end{figure}
		
		\begin{figure}[t]
		\includegraphics[width=0.49\textwidth]{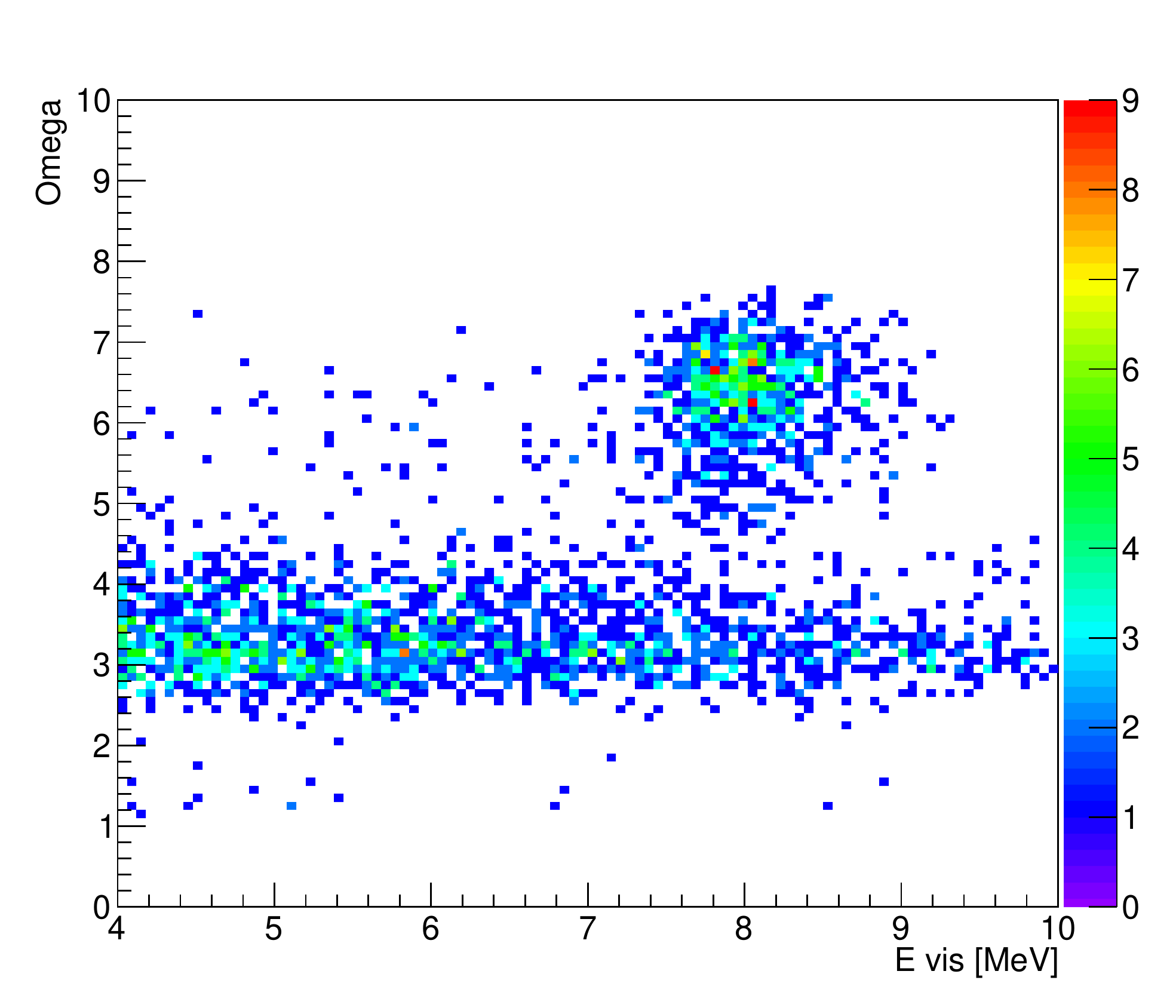}
		\hfill
		\includegraphics[width=0.49\textwidth]{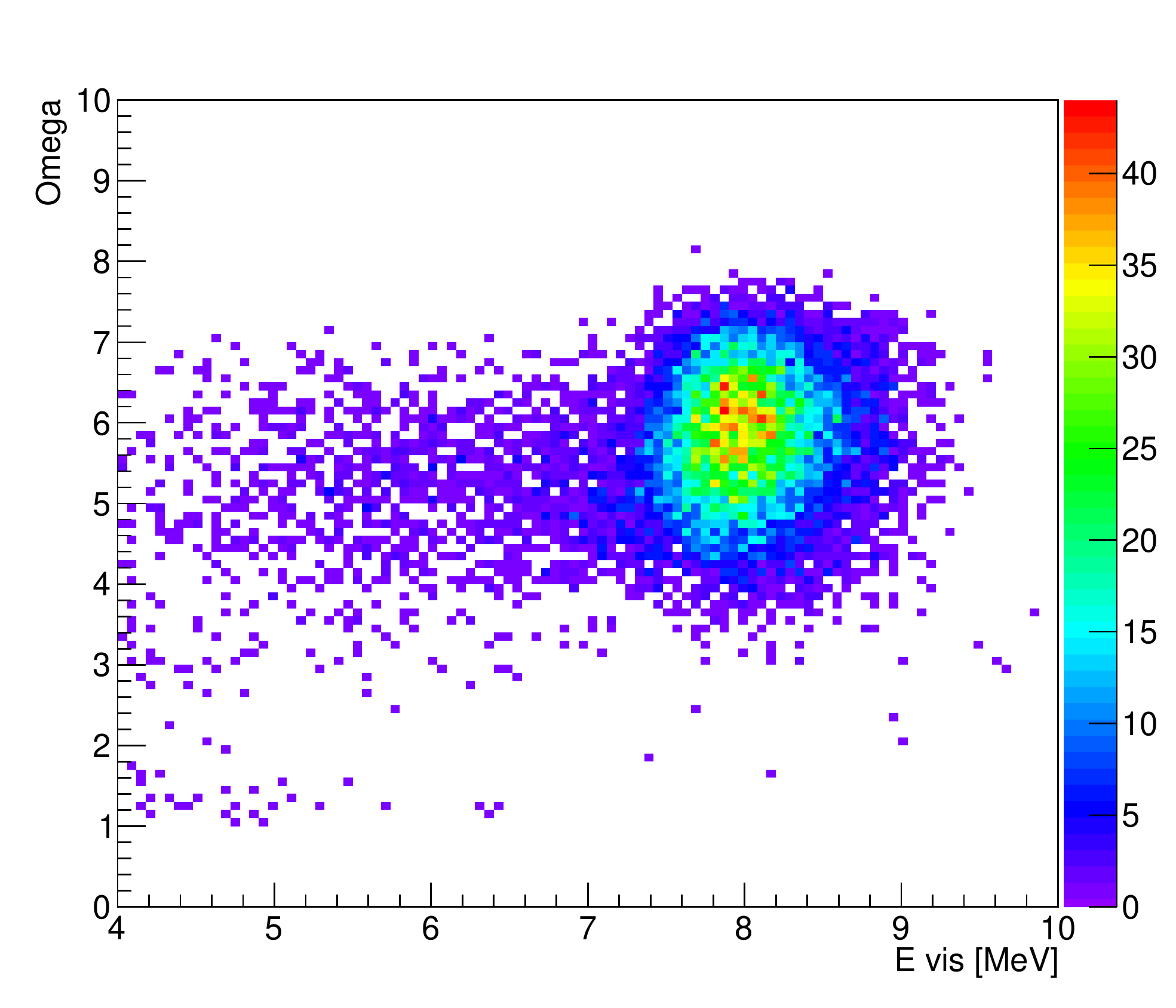}
		\caption{The discriminator variable $\Omega$ for \textit{delayed} events with a reconstructed vertex
		position inside and outside a sphere of radius $\Delta R<50$~cm around the point $z=75$~cm
		(shown as a red circle in Figure~\ref{4b}).
		\emph{(Left)} Distribution of discriminator values for delayed events within the sphere.\label{4e}
		\emph{(Right)} Distribution of discriminator values for delayed events outside of the sphere.\label{4f}}
		\end{figure}

The left panel of Figure~\ref{img:SMvertices} shows the reconstructed event vertices for the prompt events of this IBD candidate sample. Since the selection criteria are based on a delayed neutron capture on a Gadolinium nucleus, the events are almost exclusively located in the Target volume. Stopping muons appear as a large population of events in the upper part of the Target within a sphere of $\Delta R<50$~cm around the point $z=75$~cm (circled in red in the plot). Their reconstructed vertex is biased towards this position as a result of the blinding of scintillation light. The corresponding $\Omega$ values are shown in the right panel. In addition to the expected cluster of IBD events\footnote{The triangular bulge towards lower values is due to ortho-positronium formation, as explained in more detail in the following section.} they exhibit a band of lower values that are in agreement with GC events, indicating a mismatch between vertex and pulse shape information. This band is only present due to the relaxed selection cuts used in this paper and does not appear not in the official selection~\cite{DC3}.

		\begin{figure}\begin{center}
		\includegraphics[width=0.6\textwidth]{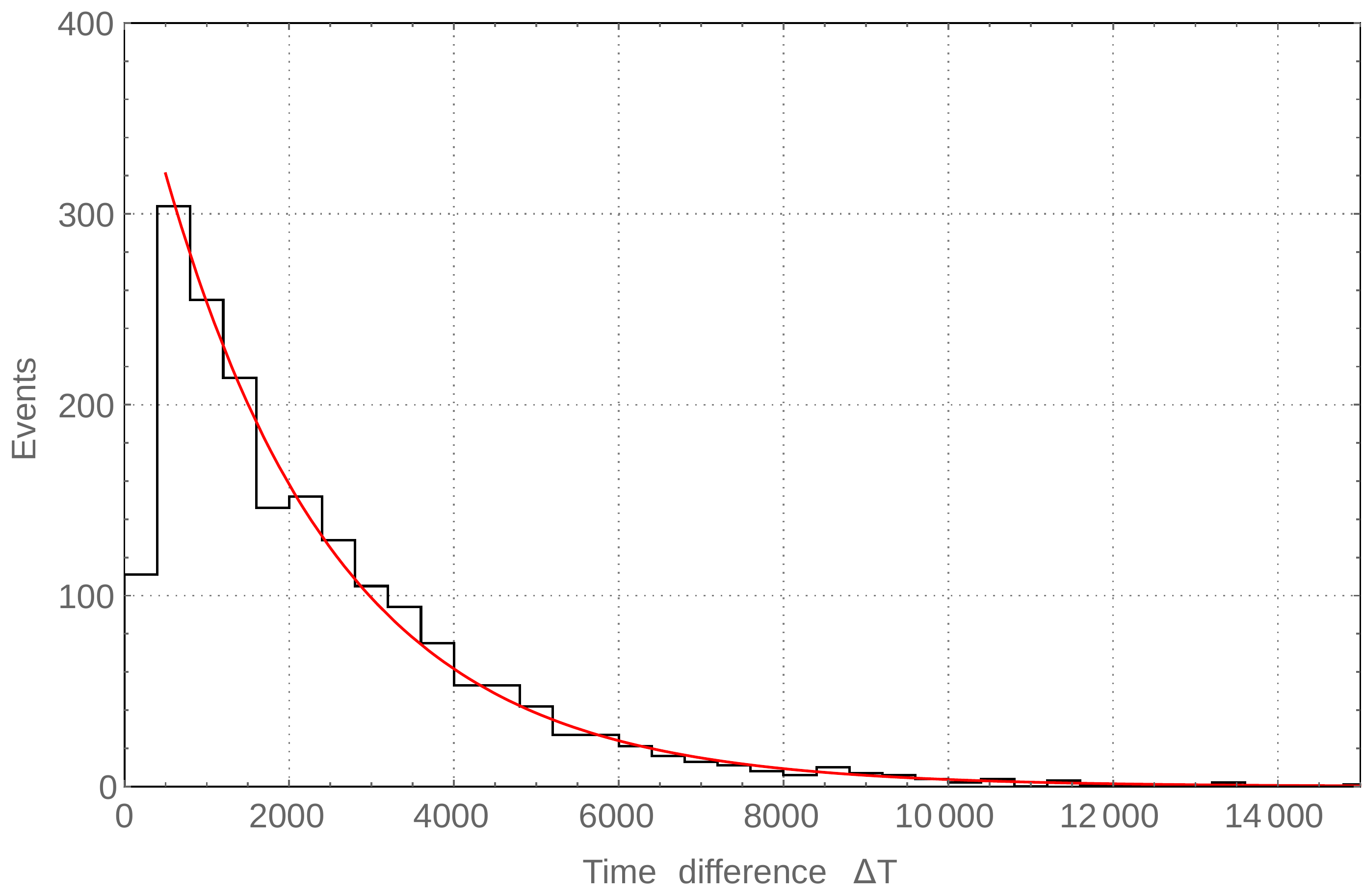}
		\caption{Distribution of time difference $\Delta T$ between the prompt and the delayed event for stopping muon
		candidates selected within the sphere of $\Delta R<50$~cm around the point $z=75$~cm and with $\Omega<4$.
		The time constant of the fit is $2.120\pm 0.052$~{\u}s in agreement with the muon lifetime.}
		\label{img:SMdeltaT}
		\end{center}\end{figure}

It is enlightening to look individually at the events within and outside of this sphere. The left panel of Figure~\ref{4c} shows the $\Omega$ values of events reconstructed within this sphere. Apart from a population of IBD events with $\Omega\gtrsim5$ it prominently features the band of GC-like values. This band vanishes for events outside the sphere, as seen in the right panel. This indicates that the events reconstructed inside the sphere are a mixture of stopping muon events and genuine IBD events, while the events reconstructed on the outside are essentially free of stopping muons.

All current selection criteria for stopping muons in Double Chooz only make use of information about the delayed event (to avoid a potential modification of the prompt energy spectrum), so it is interesting to also see the behavior of the discriminator for the delayed events. Figure~\ref{4e} shows that the corresponding delayed events --- in this case the Michel electrons of the muon decay --- exhibit a similar behavior to the prompt events, providing further evidence for stopping muons.\\
\\
The observed behavior makes it possible to identify stopping muons with a combined cut on the reconstructed vertex and the SSD discriminator value: if an event is reconstructed in the Target volume (within the spherical region), but has an $\Omega$ value consistent with GC events, it is likely a stopping muon. Figure~\ref{img:SMdeltaT} shows the distribution of the time difference $\Delta T$ between the prompt and delayed events for the events selected by these two cut conditions. A fit to this distribution with a single exponential decay function yielded a time constant of $2.120\pm 0.052$~{\u}s, in perfect agreement with the muon lifetime ($\tau_\mu = 2.197$~{\u}s). This is a confirmation that the selected events are indeed stopping muons.

\paragraph{Ortho-positronium}
Positrons from the IBD reaction usually deposit their energy in the scintillator and subsequently annihilate with an electron, and both processes contribute to the scintillation signal. The resulting pulse shape is a superposition of these two signals. But there is also a certain probability that the positron combines with an electron after its thermalization and forms ortho-positronium (o-Ps), which exists for 3.7~ns nanoseconds in the Double Chooz scintillator before it annihilates. If this happens there is a short delay between the positron energy deposition signal and the annihilation signal, which leads to a broadening of the resulting pulse shape. If the delay is large enough, two separate peaks may be identified. Ortho-positronium events can already be identified in Double Chooz with a dedicated technique (``o-Ps tagging'') described in Ref.~\cite{oPs}.\\
\\
The influence of ortho-positronium formation on $\Omega$ can be seen in the right panel of Figure~\ref{4d} from the section on stopping muons. The large population of IBD events shows a triangular bulge towards lower values. While $\Omega$ itself is not energy dependent, the \textit{visibility} of o-Ps formation with $\Omega$ depends effectively on the energy. This is because o-Ps formation leads to a stronger relative distortion of the pulse shape when the positron energy is low. The energy released by the annihilation is always 1.022~MeV, so it has a dominant influence on the pulse shape as long as the total event energy is below 2~MeV. With increasing event energies the pulse shape becomes more and more dominated by the positron energy deposition and the relative influence of the two gammas becomes gradually lower.

When o-Ps formation occurs the events tend to have lower $\Omega$ values than otherwise, since the resulting broader pulseshape gives rise to a narrower Fourier power spectrum. A cross-check with the o-Ps tagging method showed that these events with low $\Omega$ are indeed identified as o-Ps events with a comparably large time delay between the positron energy deposition and the annihilation signal. This time delay can be arbitrarily small. Since $\Omega$ increases with decreasing delay, this causes the o-Ps population to blend smoothly into the population of IBD events without o-Ps formation. Because of this the o-Ps events are not readily separable from the events without positronium formation with an SSD-based discriminator alone.


It shall be noted that the o-Ps tagging algorithm was specifically developed for the task of identifying o-Ps events and is actively searching for signatures of o-Ps in the pulse shapes. This way it can achieve an event-by-event tagging of o-Ps formation. In contrast, the SSD-based classifier rather makes a qualitative statement about the shape of the pulse as a whole. Nevertheless, the fact that the SSD approach is sensitive to o-Ps events is considered a demonstration of its particle identification potential.

\begin{figure}\begin{center}
		\includegraphics[width=0.48\textwidth]{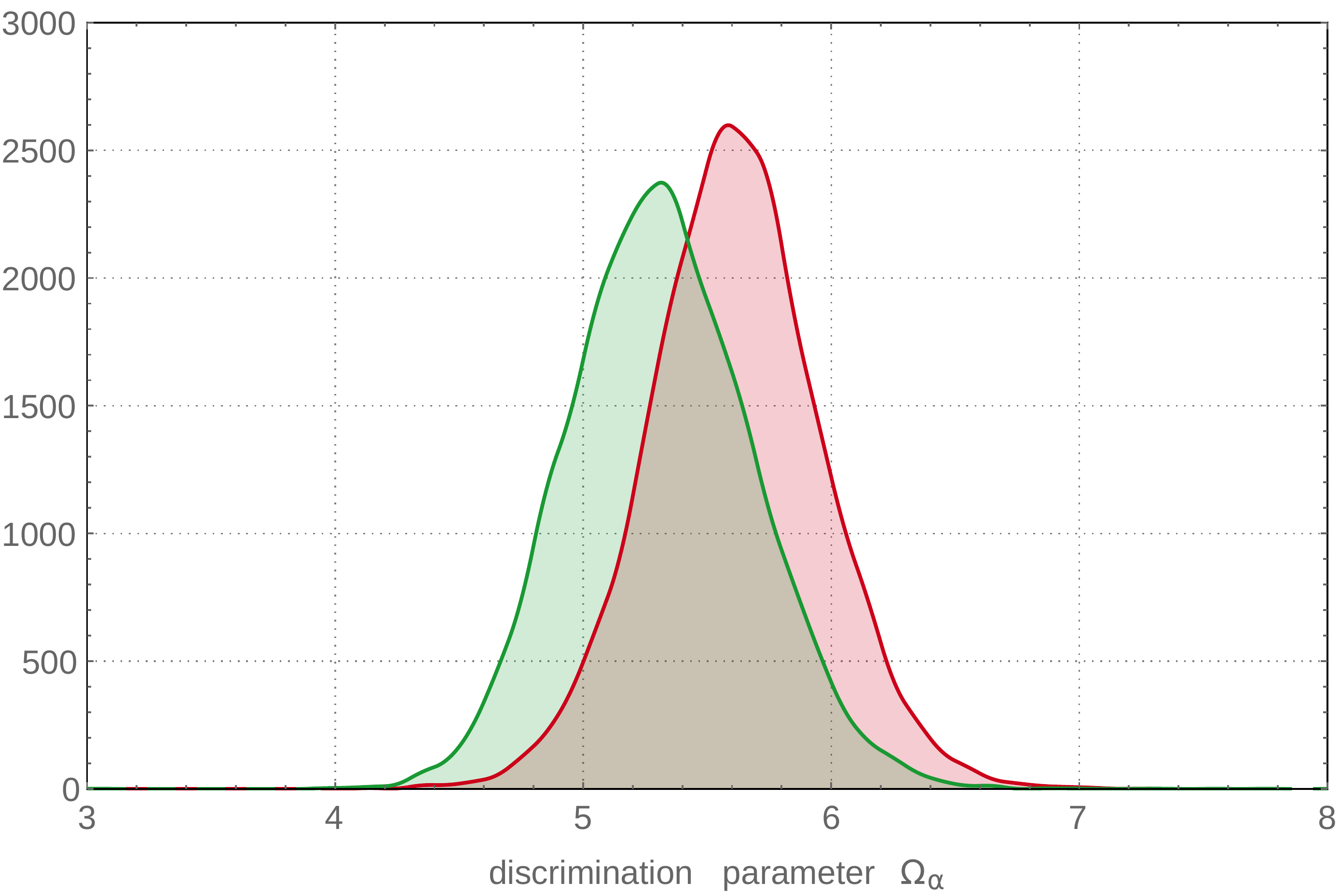}
		\caption{Distribution of the optimized spectral shape-based discriminator $\Omega_\alpha$
		for $^{212}$BiPo events in the Target volume.
		The prompt electrons (red) have a slightly higher mean value than the delayed alpha particles (green).}
		\label{img:BiPo}
		\end{center}\end{figure}

\paragraph{Alpha particles}
Alpha particles have different scintillation pulse shapes than electrons due to their distinct energy loss mechanisms and the resulting ratios of excited singlet and triplet states in the scintillator. A suitable sample of alpha and electron events can be obtained from the decay of radioactive Bismuth isotopes, which enter the detector as decay products of Radon. The isotope $^{212}$Bi undergoes beta decay and its daughter $^{212}$Po subsequently decays via alpha emission with a half-life of 299~ns, producing a coincidence signal in the detector. These events have been selected according to the criteria described in Ref.~\cite{Martin} and were restricted to the Target volume for this study. The selection cuts were:
		(i)~a visible energy of the prompt event satisfies $E_\mathrm{vis}\in[0.5,3]$~MeV,
		(ii)~a visible energy of the delayed event satisfies $E_\mathrm{vis}\in[0.35,1.2]$~MeV\footnote{The alpha energy of the
		$^{212}$Po decay is 8.95~MeV, but due to strong quenching of the scintillation light yield for alpha particles the visible
		energy lies in a much lower range.},
		(iii)~a time coincidence between the prompt and delayed event of $\Delta T \in [0.5, 5]$~{\u}s, and
		(iv)~a distance between the prompt and delayed reconstructed vertices of $\Delta R < 50$~cm.
Due to the extremely short coincidence window the sample is very clean of backgrounds events. Figure~\ref{img:BiPo} shows the distribution of a classifier $\Omega_\alpha$ optimized for alpha particles (green) and electrons (red) from the prompt and delayed events of a $^{212}$Bi-Po sample. 

Even though the two distributions overlap significantly, there is a clear shift between the mean values of the two particle species. As expected, alpha particles tend to have lower $\Omega_\alpha$ values, as their pulse shapes are broader due to a larger contribution of triplet states, confirming the sensitivity of the spectrum-based classifier to particle-induced differences in the scintillation pulse shapes. The separation between the two distributions is presently not large enough for an efficient identification of alpha background, but it may be increased with nonlinear optimization techniques and further improvements (see Section~\ref{sec:limitations}).

\paragraph{Electron/positron discrimination}
An efficient identification of positron events by means of their pulse shapes would be of highest interest for reactor neutrino experiments. The positrons created by the interaction of antineutrinos with protons are a unique feature of IBD events; no relevant backgrounds produce positrons in this energy range. An efficient electron/positron-discrimination on an event-by-event basis could thus render an IBD-sample virtually background-free, which would significantly reduce the uncertainties in $\theta_{13}$ analyses.

Most notably, an electron/positron discrimination would be able to identify the cosmogenic $\beta^-$-n emitters $^9$Li and $^8$He by their prompt electrons. These isotopes are mainly created by muon spallation reactions in the detector~\cite{MuonPaper}. They undergo $\beta^-$-decay immediately followed by the emission of a neutron, which can be captured on Gadolinium or Hydrogen. This way they produce a coincidence signature in the detector that is currently indistinguishable from genuine IBD events, making them the most dangerous correlated backgrounds in $\theta_{13}$ analyses. In Double Chooz they are statistically treated with a likelihood technique that uses the distance of an event to a preceding muon track and the number of neutron candidates following that muon~\cite{DC3}. But if the electron of the decay of $^9$Li and $^8$He could be distinguished from positrons via the event pulse shapes, this information would provide a possibility to reject this background event-wise.\\
\\
The differences in the energy loss function $\left<\mathrm{d}E/\mathrm{d}x\right>$ between electrons and positrons are too small to be detected, but the two particles should still exhibit pulse shapes differences because of the annihilation gammas in the positron signal. This was investigated with a sample of electrons from $^{12}$B events and positrons from a Gadolinium selection.
 To guarantee a fair comparison between these event classes the energy for both samples was restricted to $E_\mathrm{vis} \in [4.8,10]$~MeV in this study. In this energy window the samples are very pure and the influence of o-Ps formation is negligible. 
 
\begin{figure}\begin{center}
		\includegraphics[width=0.49\textwidth]{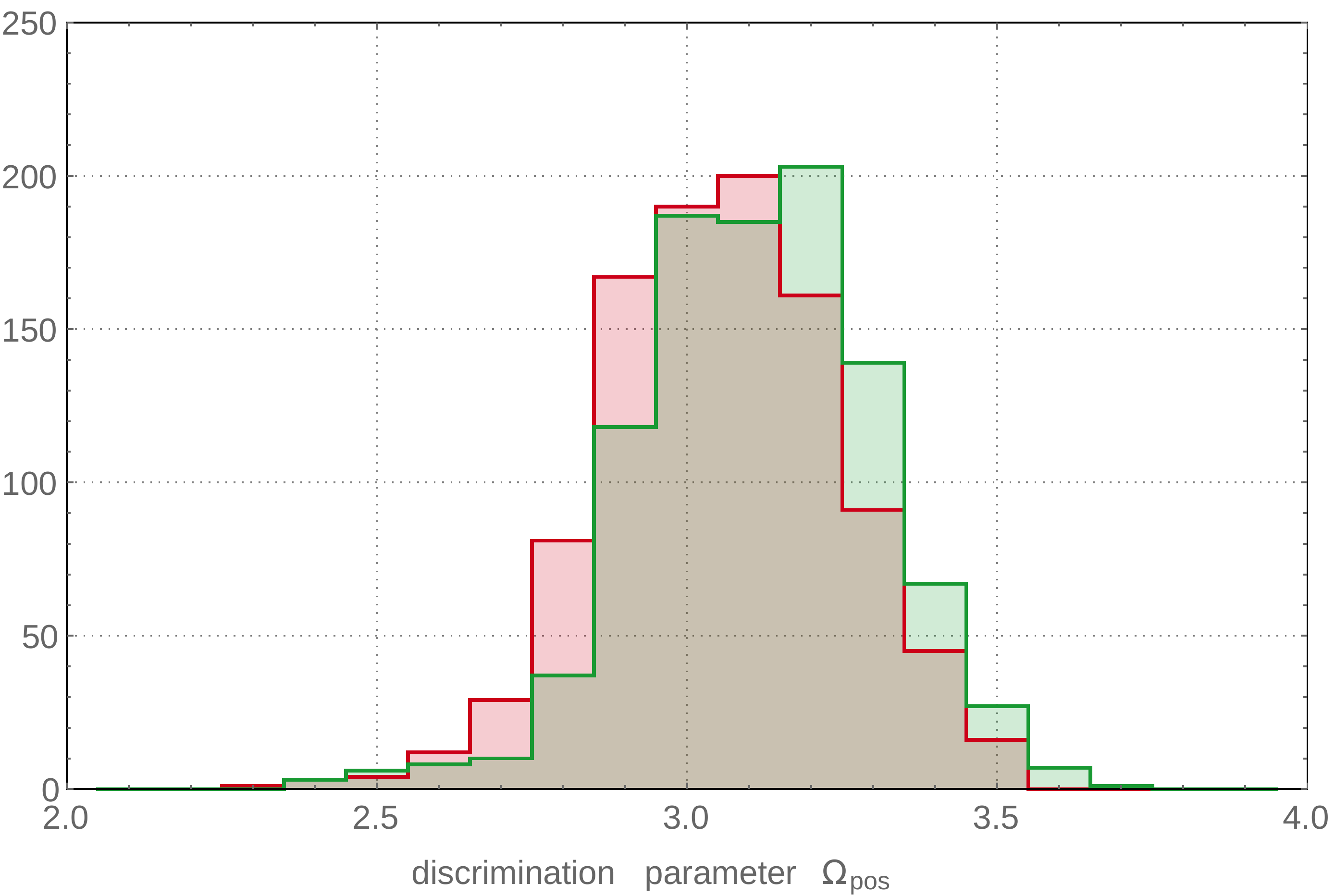}
		\hfill
		\includegraphics[width=0.49\textwidth]{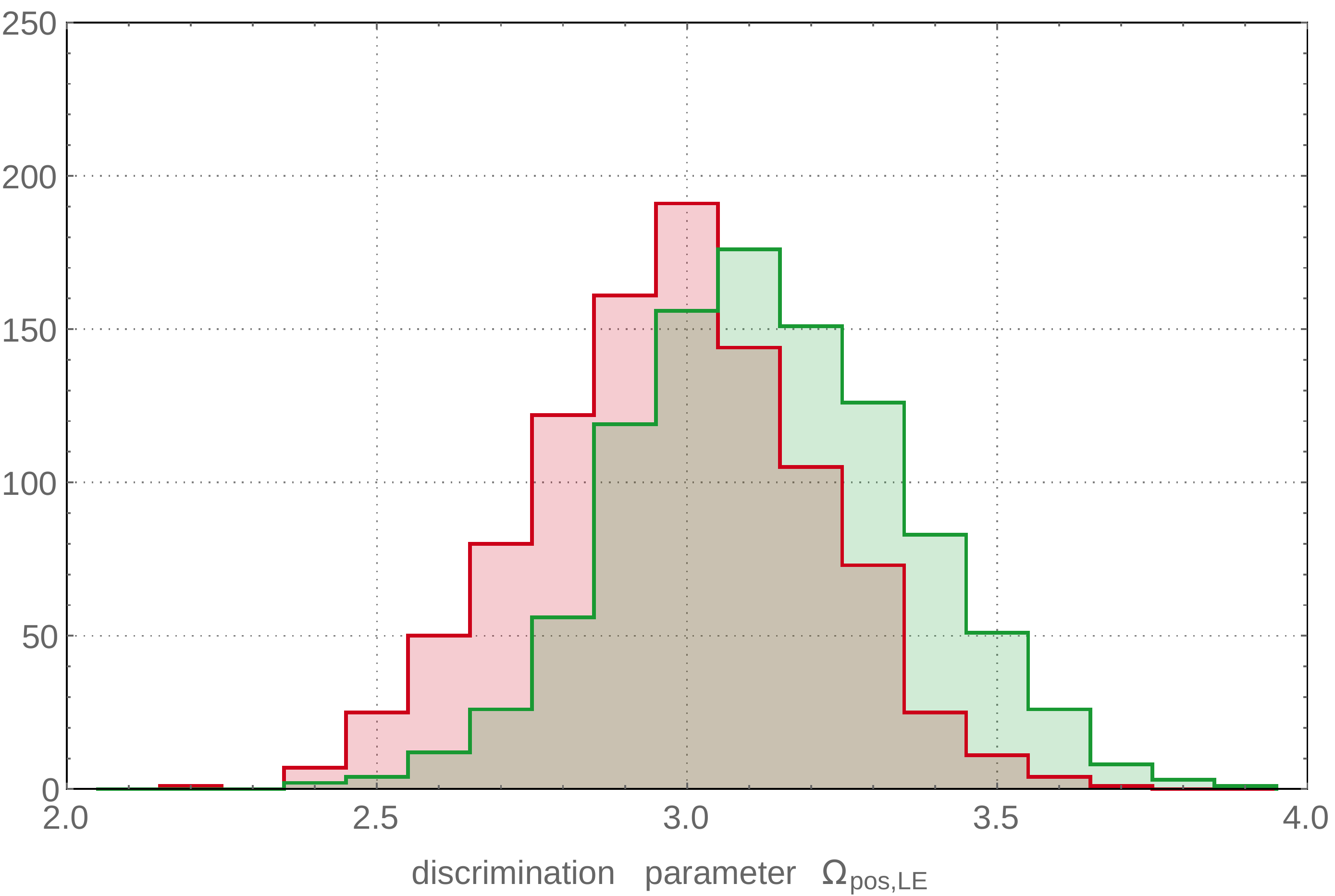}
		\caption{Distribution of the optimized discriminator for positrons (red) and electrons (green) in the Target volume. Each
		population contains 1000 events.
		\emph{(Left)} Events in the energy range of \mbox{$E_\mathrm{vis} \in [4.8, 10]$~MeV}.
		\emph{(Right)} Events in the energy range of \mbox{$E_\mathrm{vis} \in [1, 2]$~MeV} (a small amount of potential o-Ps
		candidates are removed from this positron sample).
		\label{img:ElectronPositron}}
		\end{center}\end{figure}

Figure~\ref{img:ElectronPositron} shows an optimized classification parameter for two equally-sized samples of electrons and positrons. Both distributions mostly overlap, but a slight shift in the mean values is visible. This is caused by the delayed energy deposition of the annihilation gammas in the case of positrons, slightly broadening the pulse shape and leading to somewhat lower values of the discriminator. This confirms that there are indeed some characteristic pulse shape differences between electrons and positrons that could in principle be exploited. Due to the large overlap though, there is no appreciable separation capability with a linearly optimized discriminator according to equation~\ref{eq:OmegaOpt}.

However, at lower event energies the differences increase, since the delayed annihilation gammas carry a larger fraction of the visible energy. At $E_\mathrm{vis}=2$~MeV, for instance, the deposited energy is shared equally between the energy loss of the positron and of the annihilation gammas. The resulting pulse shape gets broader and consequently the separability of the distributions increases. This was tested with low-energy events. The prompt event of the $^{212}$Bi-Po selection described in the previous paragraph provides a sample of low-energy electrons, which were compared to positrons from the same Gadolinium selection in the energy range of \mbox{$E_\mathrm{vis} \in [1, 2]$~MeV}. A small number of events with a high probability of o-Ps formation were removed. The performance of classifier optimized for two equally-sized samples of electrons and positrons at these energies is shown in the right panel of Figure~\ref{img:ElectronPositron}. Due to the reduced number of scintillation photons at lower energies the statistical variation of the pulse shapes is larger. For this reason the distributions of the classifier are wider than at higher energies, but the separation increased. At classifier values above 3.35 about 80~\% of the events are electrons, below 2.75 about 79~\% of the events are positrons. But restricting an analysis to these margins of the distribution would exclude too big a part of the neutrinos to be useful in its current state. 

The separation may improve with the possible future enhancements of the SSD technique presented in the next section. It is also worth mentioning that a separation between electrons and positrons was not expected in a large-scale liquid scintillator detector and even a small separation would be an important achievement that could find interesting applications in the future.

\section{SSD in the context of Double Chooz}\label{sec:limitations}
The Double Chooz experiment was not designed with a focus on pulseshape-based particle identification, so the environment in which SSD was developed is not optimized for pulseshape analyses. But the previous sections showed that differences between the scintillation pulse shapes of different particles are still contained in the recorded signals and that the SSD technique is sensitive to these subtle differences. Although in Double Chooz the separability is currently too small to be useful in practice, this could lead to an SSD-based particle identification in the future. This section considers the experimental factors that currently limit the performance of the technique for PID and presents potential solutions.\\
\\
The most important limitation of the separation capability of the discriminator comes from the statistical nature of the scintillation signal. When the scintillator is excited by an ionizing particle, optical photons are emitted according to a probability distribution with a specific time constant. The statistical fluctuations in the number of photons and their emission time cause a variation of the signal shape, which is effectively equivalent to a noise component (shot noise). In addition, the number of secondary electrons per PMT hit also varies statistically, which further impacts the resulting signal shape.

These stochastic noise contributions are translated into the Fourier domain and are by far the biggest source of uncertainty in the power spectrum. They account for the largest part of the RMS of the classifier and consequently also create most of the overlap between different event categories (as seen, for example, in Figures~\ref{img:BiPo} and~\ref{img:ElectronPositron}).

These kinds of statistical fluctuations are unavoidable in a detector setup with scintillators and PMTs, but could be mitigated by larger photon statistics. In a detector with a higher light level per MeV the effect of these fluctuations would be reduced and the performance of SSD is expected to increase.\\
\\
Other influences depend to a larger extent on the design and hardware choices of the experiment. One such effect would be an additive white Gaussian noise (AWGN) component induced by detector electronics and/or the readout system (e.g. fluctuations in the signal baseline). This type of noise has constant spectral density and affects all coefficients. As such, it is also incorporated into the classifier and contributes to its RMS. It is the main reason why coefficients beyond $j\approx20$ are not taken into account in the definition of the discriminator (see Figure~\ref{img:spectrum}). A lower amount of AWGN in the time domain would decrease the overall noise level in the spectrum and the lower impact on the classifier.

Another noise contribution caused by the electronics is quantization noise. The DAQ digitizes the PMT signals into discrete values with 8-bit vertical resolution. The continuous values of the photocurrent are rounded to the nearest integer value, leading to slight differences between the actual and digitized values, which is again equivalent to a noise component. For large pulses the influence of this effect is negligible, but very small pulses can suffer considerably from this effect, especially single-photoelectron pulses in individual PMTs. This kind of noise is to some extent propagated to the classifier.\\
\\
The classifier can also be affected by truncation of the PMT signals. The Double Chooz readout window is 256~ns wide and pulses of average height start approximately in the middle. While smaller pulses usually terminate before the end of the readout window, the tails of larger pulses can extend beyond the readout window so that a part is cut off. Such a truncation in the time domain is equivalent to a convolution of the Fourier spectrum with functions of the type \mbox{$\frac{a}{x} \sin\!\left(\frac{a}{x}\right)$} with different widths $a$, effectively distorting the spectra. This phenomenon can be avoided by using larger windows to accomodate all relevant pulses.\\
\\
Except for the statistical fluctuations these undesired effects on the SSD can be reduced or avoided with appropriate hardware, which might be aspects to consider in future experiments. Nevertheless, there are several potential prospects how existing experiments like Double Chooz might still be able to further increase the performance of the SSD technique and achieve a more efficient identification of particles.

One approach that is being studied is to attempt a reconstruction of the original SPE signals from fits to the recorded PMT pulses~\cite{Adrien}. This would yield smooth and noiseless signals that can be used as a new time-domain input for the SSD. This method would eliminate most of the influences presented in the previous paragraphs. In particular, electronic noise, quantization noise and statistical fluctuations originating from the PMTs are removed from the signal. In addition, the continuous output also increases the effective time resolution, leading also to a more detailed power spectrum.

Another approach aims for a better exploitation of the power spectra. The current discriminator defined in equation~\eqref{eq:OmegaOpt} is a linear combination of the spectral coefficients. A nonlinear classifier could potentially obtain more information from the spectra and achieve a better separation of event classes. This could be achieved for example with multilayer artificial neural network (ANN) that takes the relevant spectral components as inputs. An adequately designed and trained ANN might be able to find nonlinear decision boundaries that are more precise than the linear boundaries created by the current linear combination of spectral coefficients.

\section{Conclusions}\label{sec:conclusions}
The novel method of spectral shape discrimination presented herein has been demonstrated to provide an event-wise and energy-independent event discrimination in large liquid scintillation detectors. Despite the lack of an \textit{a priori} optimization of the Double Chooz detectors for pulse shape-based event identification, SSD achieved an unprecedented performance, as demonstrated by the identification of the interaction volume, instrumental light noise, stopping muon events, ortho-positronium formation, and its sensitivity to the particle type. The technique is made possible on the fast FADC electronics readout (including fast photodetectors) and the carefully designed liquid scintillator used in the experiment, and may achieve particle identification upon further optimization.

Beyond Double Chooz, we expect SSD to become a valuable handle for further background rejection in (current or future) liquid scintillator detectors with optimized designs and potentially better instrumentation. The performance, however, depends very much on the actual detector design, so it would be futile to attempt any estimates at this point.

Within Double Chooz, the SSD technique will likely receive significant benefits from further exploitation of the FADC, which aims to reconstruct single-photoelectron signals from the waveforms and is being developed within DC. This method would provide high-quality input signals for SSD to achieve maximal performance with the given detector hardware. Further improvements are possible and have been highlighted in the discussion of this paper.

\acknowledgments
We thank the French electricity company EDF; the European fund FEDER; the Région de Grand Est; the Département des Ardennes; and the Communauté de Communes Ardenne Rives de Meuse. We acknowledge the support of the CEA, CNRS/IN2P3, the computer centre CCIN2P3, and UnivEarthS Labex program of Sorbonne Paris Cité in France (ANR-10-LABX-0023 and ANR-11-IDEX-0005-02); the Ministry of Education, Culture, Sports, Science and Technology of Japan (MEXT) and the Japan Society for the Promotion of Science (JSPS); the Department of Energy and the National Science Foundation of the United States; U.S. Department of Energy Award DE-NA0000979 through the Nuclear Science and Security Consortium; the Ministerio de Economía y Competitividad (MINECO) of Spain; the Max Planck Gesellschaft, and the Deutsche Forschungsgemeinschaft DFG, the Transregional Collaborative Research Center TR27, the excellence cluster ``Origin and Structure of the Universe'', and the Maier-Leibnitz-Laboratorium Garching in Germany; the Russian Academy of Science, the Kurchatov Institute and RFBR (the Russian Foundation for Basic Research); the Brazilian Ministry of Science, Technology and Innovation (MCTI), the Financiadora de Estudos e Projetos (FINEP), the Conselho Nacional de Desenvolvimento Científico e Tecnológico (CNPq), the São Paulo Research Foundation (FAPESP), and the Brazilian Network for High Energy Physics (RENAFAE) in Brazil. Stefan Wagner would also like to thank Hiroshi Nunokawa and the Pontifícia Universidade Católica do Rio de Janeiro (PUC-Rio) for their support.

\end{document}

%% file: DCAuthors.tex
\author[a,1]{T.~Abrah\~ao,}
\author[b]{H.~Almazan,}
\author[a]{J.C.~dos Anjos,}
\author[c]{S.~Appel,}
\author[d]{I.~Bekman,}
\author[e,f]{T.J.C.~Bezerra,}
\author[g]{L.~Bezrukov,}
\author[h]{E.~Blucher,}
\author[i]{T.~Brugi\`ere,}
\author[b]{C.~Buck,}
% ...

\affiliation[a]{Centro Brasileiro de Pesquisas F\'isicas, 22290-180 Rio de Janeiro, Brazil}
\affiliation[b]{Max-Planck-Institut f\"ur Kernphysik, 69117 Heidelberg, Germany}
\affiliation[c]{Physik Department, Technische Universit\"at M\"unchen, 85748 Garching, Germany}
\affiliation[d]{III. Physikalisches Institut, RWTH Aachen University, 52056 Aachen, Germany}
\affiliation[e]{SUBATECH, CNRS/IN2P3, Universit\'e de Nantes, Ecole des Mines de Nantes, 44307 Nantes, France}
\affiliation[f]{Research Center for Neutrino Science, Tohoku University, 980-8578 Sendai, Japan}
\affiliation[g]{Institute of Nuclear Research of the Russian Academy of Sciences, 117312 Moscow, Russia}
\affiliation[h]{The Enrico Fermi Institute, The University of Chicago, Chicago, IL 60637, USA}
\affiliation[i]{IPHC, Universit\'e de Strasbourg, CNRS/IN2P3, 67037 Strasbourg, France}
% ...

\emailAdd{stefan.wagner@apc.in2p3.fr}